\documentclass[structabs]{aa}  

\usepackage{xspace,amsmath}
\usepackage{color,epsfig,amssymb,lscape}
\usepackage{util}
\usepackage{array,colortbl}
\usepackage[colorlinks=true,linkcolor=magenta,citecolor=blue,breaklinks=true]{hyperref}
\usepackage{ifpdf}

 \newcommand{\hii}{\relax \ifmmode {\mbox H\,{\scshape ii}}\else H\,{\scshape ii}\fi}
\newcommand{\mi}{\relax \ifmmode {\mu{\mbox m}}\else $\mu$m\fi}
\newcommand{\ha}{\relax \ifmmode {\mbox H}\alpha\else H$\alpha$\fi}
\newcommand{\hb}{\relax \ifmmode {\mbox H}\beta\else H$\beta$\fi}

\newcommand{\sii}{\relax \ifmmode {\mbox S\,{\scshape ii}}\else S\,{\scshape ii}\fi}
\newcommand{\siii}{\relax \ifmmode {\mbox S\,{\scshape iii}}\else S\,{\scshape iii}\fi}
\newcommand{\nii}{\relax \ifmmode {\mbox N\,{\scshape ii}}\else N\,{\scshape ii}\fi}
\newcommand{\oi}{\relax \ifmmode {\mbox O\,{\scshape i}}\else O\,{\scshape i}\fi}
\newcommand{\oii}{\relax \ifmmode {\mbox O\,{\scshape ii}}\else O\,{\scshape ii}\fi}
\newcommand{\oiii}{\relax \ifmmode {\mbox O\,{\scshape iii}}\else O\,{\scshape iii}\fi}
\newcommand{\arcsecond}{\hbox{$^{\prime\prime}$}} 
\newcommand{\arcminute}{\hbox{$^{\prime}$}} 
\newcommand{\rdostres}{\relax \ifmmode {\,\mbox{R}}_{\rm 23}\else \,\mbox{R}$_{\rm 23}$\fi} 

\newcommand{\mic}{$\mu$m}

\newcommand{\gsim}{\hbox{\rlap{\lower.55ex\hbox{$\sim$}} \kern-.3em
\raise.4ex \hbox{$>$}}}
\newcommand{\lsim}{\hbox{\rlap{\lower.55ex\hbox{$\sim$}} \kern-.3em
\raise.4ex \hbox{$<$}}}

\usepackage{graphicx}
\begin{document}

\title{The giant H{\sc ii} region NGC~588 as a benchmark
for 2D photoionisation models}
\titlerunning{A 2D model for NGC~588}

  \author{E. P\'erez-Montero \inst{1}
  \and A.Monreal-Ibero \inst{1,2}
  \and M. Rela\~no \inst{3}
  \and J.M. V\'\i lchez \inst{1}
  \and C. Kehrig \inst{1}
   \and C. Morisset \inst{4}}

 \offprints{E. P\'erez-Montero}

  \institute{
Instituto de Astrof\'\i sica de Andaluc\' \i a - CSIC. Apdo. 3004, 18008, Granada, Spain \\
            \email{epm@iaa.es, jvm@iaa.es, kehrig@iaa.es}
       \and
	Leibniz-Institut f\"ur Astrophysik Potsdam (AIP). An der Sternwarte 16, 14482 Potsdam, Germany
	\email{}	
    	\and
        Departamento de F\'\i sica Te\'orica y del Cosmos. Universidad de Granada, Campus Fuentenueva, Granada, Spain.\\
       \email{mrelano@ugr.es}
        \and
        Instituto de Astronom\'\i a, Universidad Nacional Aut\'onoma de M\'exico, Apdo. Postal 70264, 04510, M\'ex. D. F., Mexico
	\email{chris.morisset@gmail.com}
}

\date{}

   \keywords{
galaxies: individual: M33, ISM -- ISM: abundances, dust, extinction, structure -- \hii\ regions: abundances}

%

\abstract
{}
{We use optical integral field spectroscopy 
and 8\mi\ and 24\mi\ mid-IR observations of the giant \hii\ region NGC~588 in the disc of M33
as input and constraints for two-dimensional tailor-made photoionisation models
under different geometrical approaches. We do this to
explore the spatial distribution of gas and dust in the interstellar
ionised medium surrounding multiple massive stars.}
{Two different geometrical approaches are followed for the modelling structure:
i) Each spatial element of the emitting gas is studied individually using models which
assume that the ionisation structure is complete in each element to look for azimuthal
variations across gas and dust. ii) A single model is
considered, and the two-dimensional structure of the gas and the dust are assumed to be due to
the projection of an emitting sphere onto the sky.}
{The models in both assumptions reproduce the radial profiles of \hb\ 
surface brightness, the observed number of ionising photons, and 
the strong optical emission-line relative intensities. The first approach
produces a constant-density matter-bounded thin shell of variable thickness 
and dust-to-gas ratio, while the second gives place to 
a radiation-bounded thick shell sphere of decreasing particle density.
However, the radial profile of the 8\mic/24\mic\ IR ratio, depending on the gas and dust geometry, 
only fits well when the thick-shell model is used.
The resulting dust-to-gas mass ratio, which was obtained  empirically from the derived dust mass 
using data from {\em Spitzer}, also has a better fit using the thick-shell solution. 
In both approaches, models support the 
importance of the low surface-brightness positions on the integrated spectrum of the nebula,
the chemical homogeneity, the ionisation-parameter radial decrease, and the robustness
of strong-line methods to derive the equivalent effective temperature in extended
regions.
These results must be taken with care in view of the very low extinction values
that are derived from the IR, as compared to that derived from the Balmer decrement.
Besides, the IR can be possibly contaminated with the emission from a cloud of 
diffuse gas and dust above the plane of the galaxy detected at 250~\mic\ {\em Herschel}
image.}
{} 

%

\maketitle

\section{Introduction}

High surface brightness\hii\ regions in the optical spectral range are luminous tracers
of both the physical properties and chemical abundances
of the interstellar medium (ISM) of the galaxies where they are located. 
Indeed, \hii\ regions in star-forming galaxies are one of the most widely 
objects used to find out these properties throughout the Universe.
In non-resolved \hii\ regions the optical spectra are collected by
means of integrated fibres or long-slit techniques, while integral field spectroscopy (IFS)
allows spectral measurements with 2D spatial resolution for
objects in the Local Universe.
The studies of the structure of \hii\ regions then become more
complex, and therefore more elaborated depictions become now necessary.
This opens the gate to a better understanding of the interplay between 
stars, dust, and gas appearing in different spatial positions and to assess if the
assumptions made to study integrated observations in non-resolved
objects lead to accurate determinations of their properties.

The most challenging issues that come from the study of the two-dimensional
structure of \hii\ regions appears as a consequence of the lack of
spatial uniformity in many of their properties. For instance, it
is shown by \cite{ercolano} and \cite{jamet08}
that the distribution of the ionising stars in relation to the gas alters
the ionisation structure and the electron temperature.
Other authors (e.g., \citealt{gia04}) point out the
relevance of gas density inhomogeneities, which also lead
to inhomogeneities in the other physical properties derived
from optical spectroscopy. 
\cite{marcelo} also show
how the ionisation structure of the gas depends on the
amount of available gas in different matter-bounded configurations, which
gives place to large fractions of escaping ionising photons and
then affects the observed emission-line ratios used to derive the physical
properties and chemical abundances of the gas.
 
Photoionisation models constitute powerful tools for the interpretation of
the physics involved in \hii\ regions. Under appropriate geometrical considerations, models allow us to
relate the collected observational information to both quantitative and
qualitative characterisations of the studied objects and to derive their
physical properties and chemical abundances.
Models that are 3D are the most suitable appliances to spatially disentangle
the effects of different ionising sources 
on the non-uniform surrounding gas. Nevertheless,  the 
lack of observational information about the distribution of gas and stars along
the line of vision prevents
this kind of 3D model from fitting many optical IFS data on \hii\ regions most of the time.
On the other hand, other techniques based on photoionisation models try to describe  
the observed spatial variations in ionised gaseous nebulae as a consequence of the
projection of a 3D structure on a plane. This is the case of the codes, NEBU\_3D \citep{NEBU3D} or {\sc Cloudy 3D} \citep{C3D}.
In a novel approach to reproduce IFS data,  \cite{mod595} use
1D photoionisation models to fit the optical IFS \citep{relano10}
and   8~\mic\ and 24~\mic\ mid-IR {\em Spitzer} bands properties of
the Giant \hii\ Region (G\hii R) NGC~595 in the disc of M33. In that work, 
different annuli around the ionising source are defined in the area covered 
by a mosaic of several integral field unit (IFU) pointings,
and their measured integrated properties are later fitted by the models.
These models depict a uniform metallicity across a thin shell,
whose optical and IR observed structure can be explained with
azimuthal variations (i.e., in the plane of the galaxy) 
in some of the properties of the \hii\ region, as in
the dust-to-gas ratio and the matter-bounded geometry.

Since both azimuthal variations and projection effects are expected to co-exist
as causes of the observed spatial variations throughout the distribution
of gas and dust in \hii\ regions, it is necessary to explore both
model strategies in well-known and characterised objects, as is the case of
the G\hii R NGC~588, which is also in M33.
Two-dimensional observations of NGC~588 in both optical and mid-IR 
are described in \cite{monreal11} (hereafter MI11); thus, a 2D
observational characterisation of the properties of the ionised gas and the 
hot dust is possible. Besides, previous studies on this G\hii R, which are based on
ground and spacecraft imaging in different bands from the UV up to
the NIR, were used by \cite{jamet04} to study the location and
nature of the ionising stellar population.

The aim of this work is to study the possible causes of the 
observed spatial variations across NGC~588 for both the optical and
mid-IR properties to test the solidness of the strong-line methods used
in integrated observations. This is done to derive physical properties and chemical
abundances by means of photoionisation models of the
well-studied G\hii R NGC~588.
To do so, we took two different assumptions: i) an improved version of the approach 
employed by \cite{mod595} for NGC~595, based on different models for the 
individual observed spatial elements which are used to explain the observations that are
caused by azimuthal variations of its properties throughout the field of view
and ii) a single model projected
onto the sky which is used to explain the variations as a consequence of the perspective.
In the next section,
we describe the 2D structure of NGC~588 and the data sampling of the spatial distribution for both optical and mid-IR data studied in MI11
and the derivation of the integrated dust-to-gas ratio from dust temperature
and total H{\sc i} mass.
In Section 3, we present our models and in Section 4, we discuss
our results. Finally, we summarise our results and
conclusions in Section 5.


\begin{figure*}
\begin{minipage}{180mm}
\centerline{
\psfig{figure=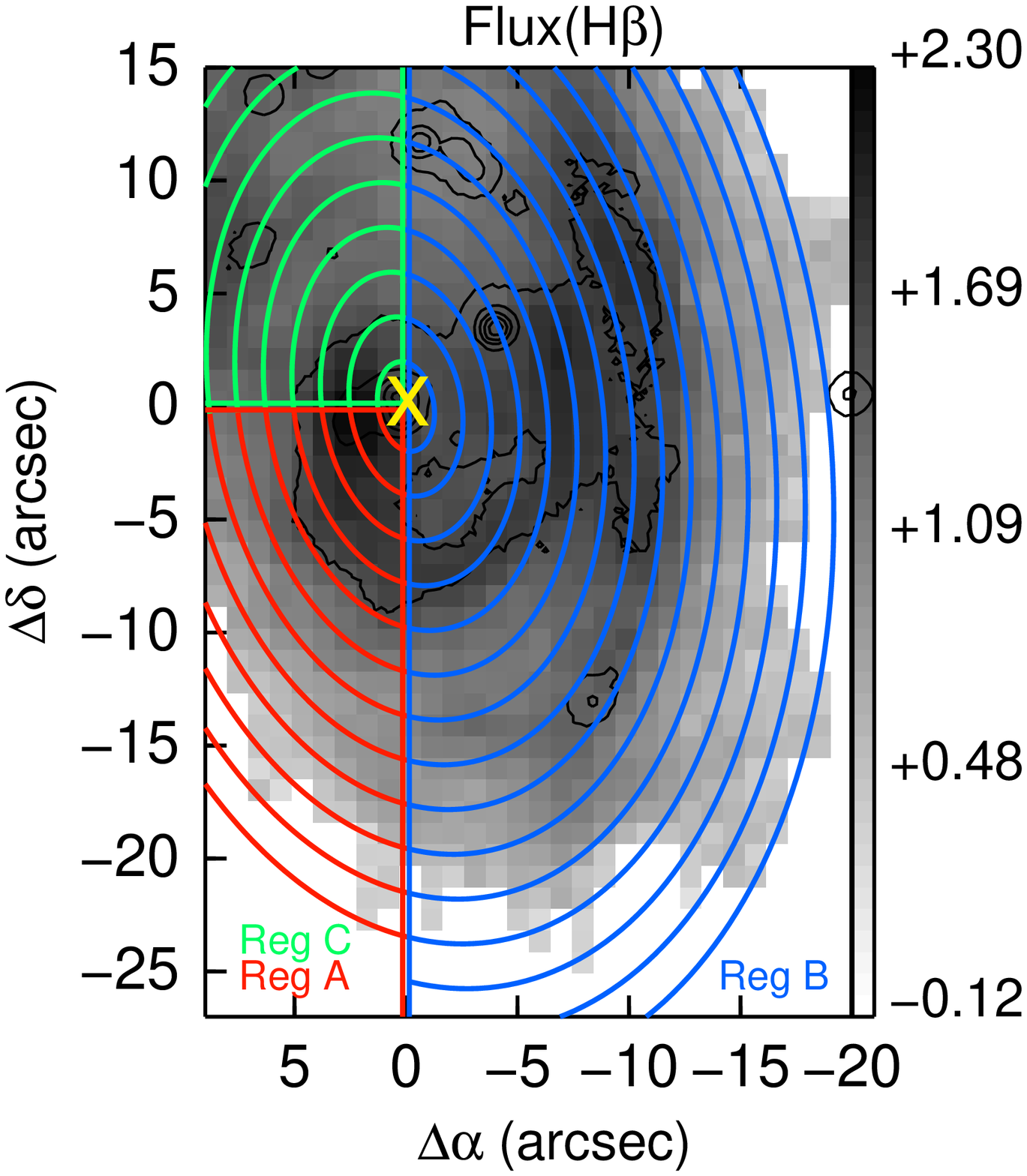,width=6cm,bb=55 30 525 565,clip=}
\psfig{figure=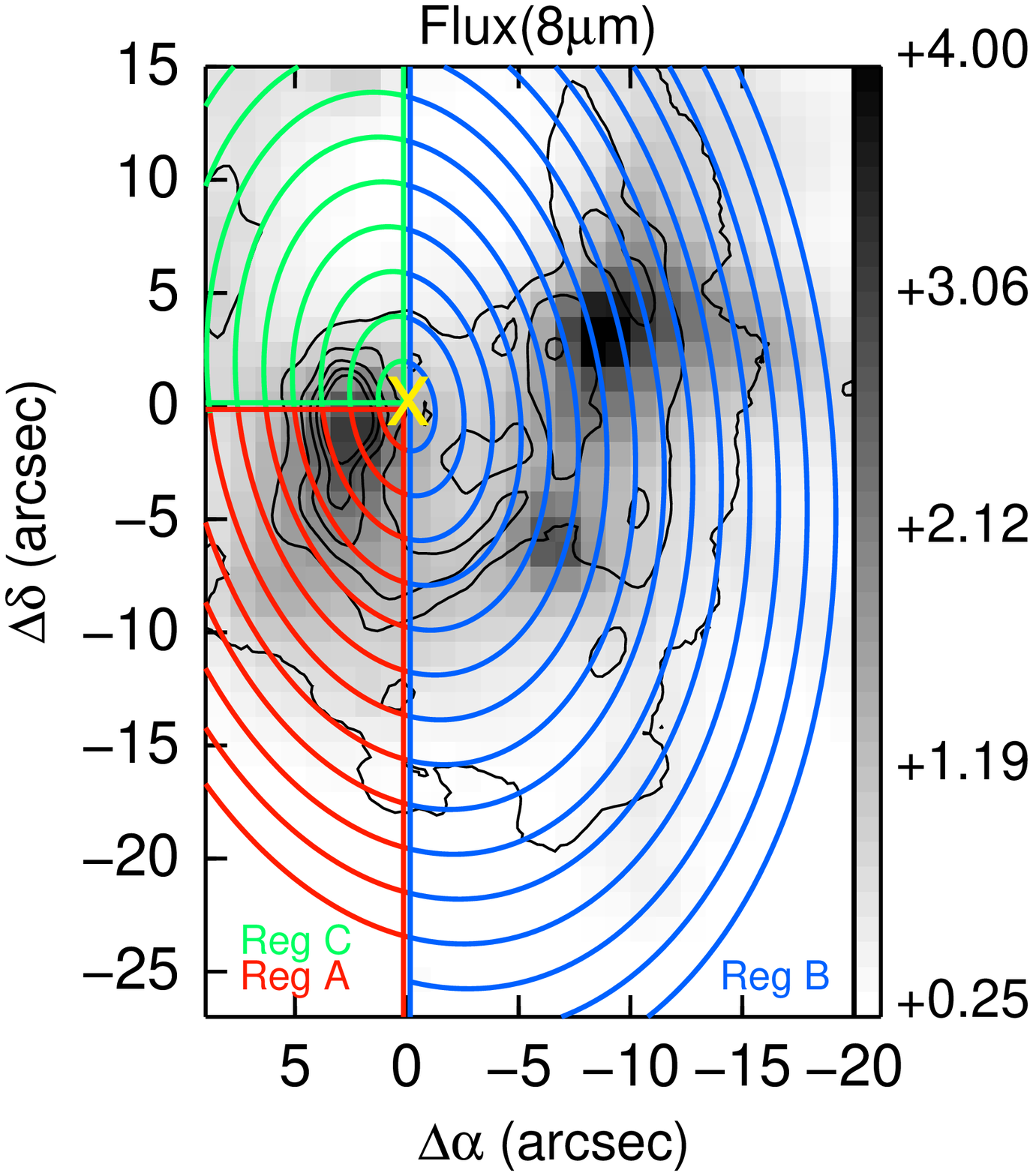,width=6cm,bb=55 30 525 565,clip=}
\psfig{figure=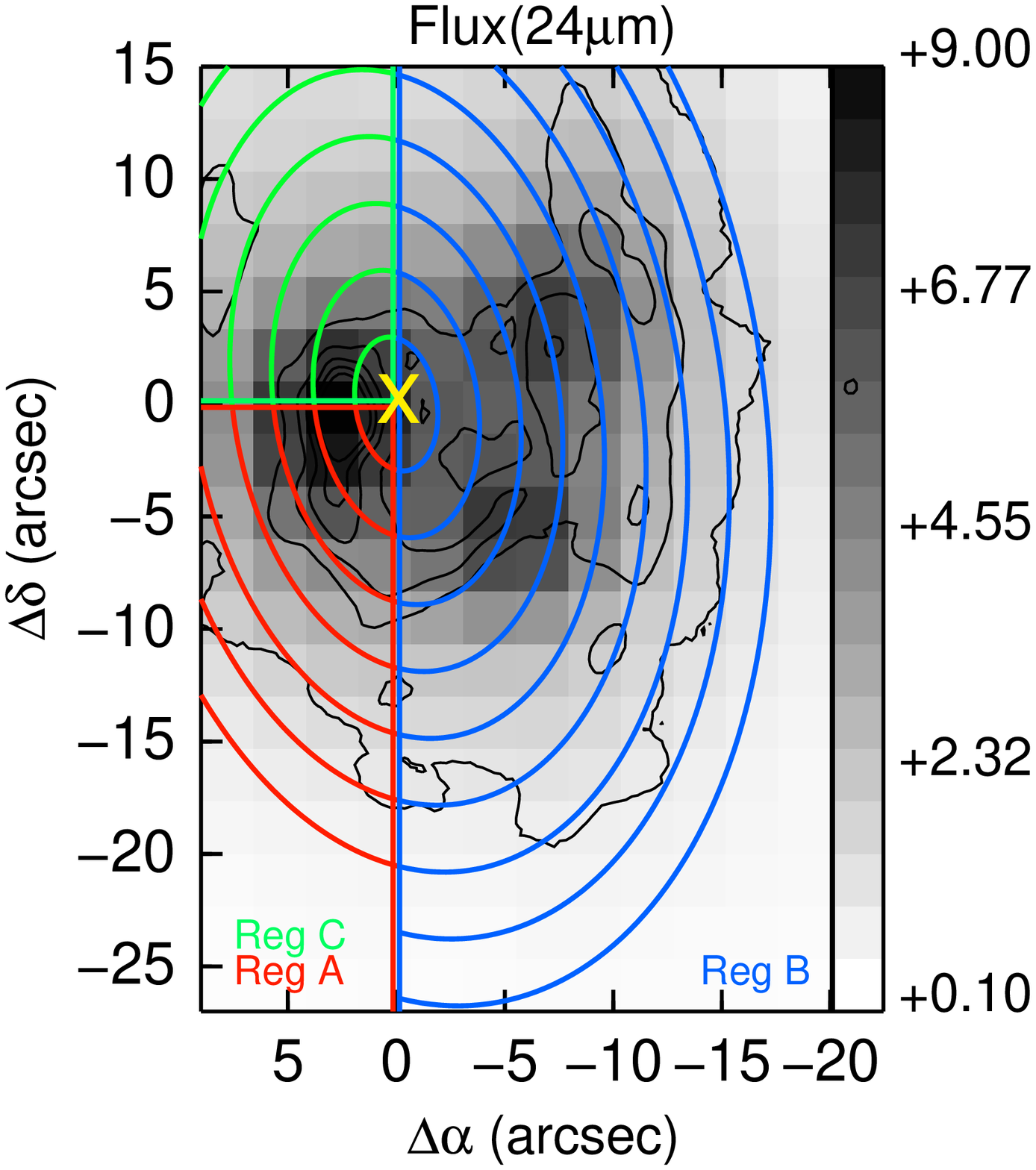,bb=55 30 525 565,width=6cm,clip=}}
\label{annuli}
\caption[]{Elliptical annular regions utilised to extract the radial 
profiles on top of the non-calibrated \hb\ (\emph{left}), 
8~$\mu$m  (\emph{middle}) and 24~$\mu$m (\emph{right}) emission 
distributions (see MI11 for more detailed plots in these
bands). Units are arbitrary in logarithmic scale.
The origin of coordinates is located at R.A. 
(J2000): 1h 32m 45.7s, DEC. (J2000): +30$^{\circ}$ 38'' 55.1' 
and is marked with a yellow "X". The three elliptical regions 
considered through the paper are labelled as "A" (\emph{red}), "B" 
(\emph{blue}), and "C" (\emph{green}) and sample the SE, W, and NE part of 
NGC~588, respectively. In all images, N points to up and E to left.}
\end{minipage}
\end{figure*} 

\section{Data description and analysis}

\subsection{Properties and 2D structure of NGC~588}

The optical observations modelled here were taken on October 9-10, 2009 
during the commissioning run of the new \emph{Potsdam Multi-Aperture 
Spectrophotometer} (PMAS, \citealt{roth}, \citealt{Roth10}) CCD on the 3.5 m telescope 
at the Calar Alto Observatory (Almer\'\i a, Spain). The PMAS was used in the lens array mode (LARR), which is 
made out of 16$\times$16 square elements, with the magnification scale 
of 1$\times$1 arcsec$^{2}$. A mosaic of six tiles was needed to 
map most of the surface of NGC~588.
The V600 grating and the 2$\times$2 binning mode were also utilised, achieving 
a $\sim$3.4~\AA\ full width half maximum spectral resolution and 
covering a spectral range of 3\,620 \AA - 6\,800 \AA. This is adequate for 
the needs of our modelling, since the main optical emission lines, 
including [\oii] $\lambda$ 3727 \AA, \hb, [\oiii] $\lambda$ 5007 \AA, \ha,
[\nii] $\lambda$6584 \AA, and [\sii] $\lambda$6717, 6731 \AA,
were observed.
Further details of the observations and data reduction can be found in 
MI11.


The IR data of NGC~588 analysed in this paper were taken from the {\em
  Spitzer} Data Archive: the 8 $\mu$m image from IRAC (Infrared Array
Camera, \citealt{fazio}) and 24 $\mu$m - 160 $\mu$m from MIPS
(Multiband Imaging Photometer, \citealt{rieke}). The spatial
resolutions of the images at 8, 24, 70, and 160 $\mu$m are
$\sim$2\arcsecond, $\sim$6\arcsecond, $\sim$18\arcsecond, and
$\sim$40\arcsecond, respectively.  The stellar contribution at 8
$\mu$m was subtracted using the emission at 3.6 $\mu$m, following the
method described in \cite{helou} and \cite{calzetti}.  The IR
observations and data reduction of 8~\mi, 24~\mi, 70~\mi, and
160~\mi\ images are described in \cite{verley}.  We also used the 250~\mi\ emission map from the {\em Herschel} telescope in this
study.  The
emission at 250~\mi\ clearly delineates the \ha\ emission of
\hii\ regions with shell morphology (see Fig.3 of \citealt{verley10}).
This map was taken as part of the ‘Open Time Key Project’ HerM33es
\citep{kramer10}.

NGC~588 has a much more complex and irregular reddening structure
than NGC~595.  Overall, it is more excited (e.g., in the brightest
\ha\ regions log([O{\sc ii}]/[O{\sc iii}]) $\sim$ 0.15 in NGC~588 and
$\sim$ 0.9 in NGC~595). This excitation difference is also evident by
inspecting the relative fluxes of other emission lines not affected by
dust depletion. \cite{rubin08} measure log([Ne{\sc ii}]/[Ne{\sc iii}])
$\sim$ 0.10 in NGC~588 and $\sim$ 0.86 in NGC~595 from {\em
  Spitzer}/IRS data.  The integrated 8\mic/24\mic\ ratio measured by
{\em Spitzer} in NGC~588 is $\sim$ 2 times higher than that in
NGC~595. Since the 8\mic\ is associated mainly with the emission of
polycyclic aromatic hydrocarbons (PAHs) formed in the
photodissociation region (PDR) and the emission at 24\mic\ owes
mainly to the small and hot dust grains mixed with the ionised gas,
this ratio can be used as an indicator of the geometry in
\hii\ regions (e.g., \citealt{bendo}).  The velocity maps in NGC~588, which are
measured by MI11, are typical for evolved \hii\ regions, where most of
the mechanical energy has already been ejected into the ISM. No
density structure from [\sii] emission lines is found.  Regarding the
brightest optical emission lines, it is found that the relative
intensity of low excitation lines, as [\oii], [\sii], or [\nii],
increase radially, in contrast to [\oiii], which decreases at a larger
distance to the ionising sources.  The different radial structure of
high- and low- excitation emission lines implies that some diagnostics
based on strong lines can present large variations across the
nebula. On the contrary, other estimators based only on low-excitation
lines, such as N2O2 or N2S2, which trace the nitrogen-to-oxygen (N/O)
abundance ratio \citep{pmc09} remain uniform within the errors across
the gas.  It is therefore  that the
ionisation parameter decreases radially and there is a chemical
homogeneity across the region found from several strong-line methods.

Considering the metal content of NGC~588, \cite{vilchez88}
derive an oxygen abundance 12+log(O/H) = 8.30 $\pm$ 0.06, which is
consistent with the radial distance of this region to the centre of M33
(e.g., \citealt{RS2008}).
Nevertheless, its log(N/O) = -1.52 $\pm$ 0.07 is lower than the expected value
with the same criterion (e.g., \citealt{magrini07}). In a later work, \cite{jamet05} derive
a slightly lower metallicity [12+log(O/H) = 8.16 $\pm$ 0.07, 0.3$\cdot$ Z$_{\odot}$] and a higher
log(N/O) [=-1.39 $\pm$ 0.09], but they argue that no confident 
estimation of these abundances can be made, since they obtain an electron
temperature of [\oiii] based on the ratio of $\lambda$ 5007 \AA\ 
and $\lambda$ 88 $\mu$m emission-lines, which are about 3000 K lower than
those derived using similar optical diagnostic ratios. They then consider
different geometrical approaches to make 1D tailor-made models 
of this region, but  electron temperatures and ionic abundances
are quite similar to those obtained from the measured optical emission lines in all cases.

\subsection{Data sampling}


NGC~588 presents a ring-like morphology in \ha\ 
(see maps in MI11 and also in \cite{cmt96}).
Therefore, the strategy to extract the spatial information is
similar to the one adopted in \citealt{mod595} for NGC~595, which is based on
elliptical concentric annuli.
However, we found  that the ionisation 
structure does not necessarily follow the same ring-like symmetry in MI11. In 
particular, the [\oiii]$\lambda$5007\AA/\hb\ line ratio 
is especially high in the SE quadrant, while 
the values of the [\nii]/\ha, [\sii]/\ha, [\oiii]/[\oii], and [\oii]/\hb\ line ratios in 
the NE quadrant correspond to larger ionisation parameters 
than those measured at similar distances to the emission peak. 
Therefore, we decided to 
model the GH{\sc ii}R by dividing it in three regions, as 
depicted in Fig. \ref{annuli}. Hereafter, we refer to the 
corresponding defined regions as A (SE quadrant), B (W half) and C 
(NE quadrant).

We divided each region in annuli, which were considered as circles onto
the plane of the galaxy but have an elliptical shape owing to the 
inclination of the disc onto the sky.
The elliptical annuli were all centred at the location of the peak 
of emission in the continuum (R.A. (J2000):
1h32m45.7s, DEC.(J2000): +30$^{\circ}$38\arcminute 55.1'').  The position of the centre and the
annuli are shown in Fig. 1 compared to the emission of the gas and dust
in different wavelengths.
The major to minor axis ratio of the
ellipse was derived using the inclination angle of the galaxy 
(i=56$^{\circ}$ for M33, \citealt{vandenBergh}), and we chose a position 
angle of 10$^{\circ}$ for the major axis, since this orientation better 
traces the shell structure of the region (see Fig. \ref{annuli}). 
In this configuration, we used rings of 2\arcsecond width to obtain the 
profiles for the elliptical regions from our IFS observations and 
8~$\mu$m image, while rings of 6\arcsecond width were utilised for the 
24~$\mu$m image. The covered radial distance varies, depending on the 
region, which can range from about 90~pc for region C to 
$\sim$150~pc for region B if we consider the distance to M33
adopted in MI11, which is of 840 kpc.

We applied masks to the data cube of our IFS observations to isolate the emission coming 
from each annulus and the integrated spectra for each of them which were produced. 
The spectra were analysed using the {\sc MPFITEXPR} algorithm 
\citep{markwardt} 
to fit the different emission lines with a Gaussian function plus a 1-degree polynomial 
function for the continuum subtraction. The final result of this procedure is a set of 
integrated fluxes for the emission lines fitted in the spectrum corresponding to each 
elliptical annulus. Flux errors were obtained as a combination of those derived in the 
profile fitting procedure and the uncertainty in the continuum subtraction 
(see \citealt{pmd03} and references therein). The IFS observations were not taken 
under photometric conditions. Since no absolute calibration was performed with the 
help of auxiliary images, all the fluxes are in relative units (see MI11 for details).

\begin{figure}
\psfig{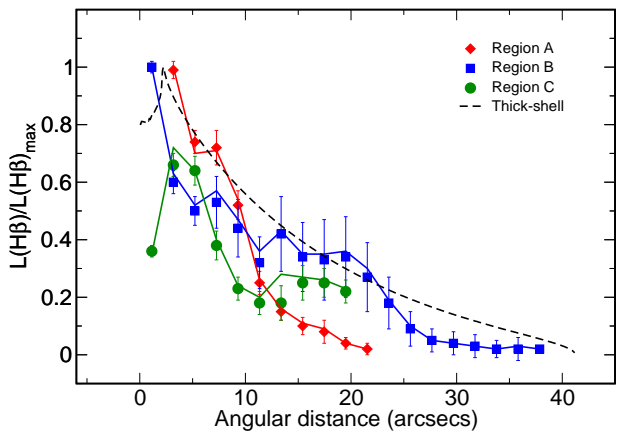}
\caption{Radial profile in observed angular units of \hb\ surface
brightness normalised to the maximum 
in the three defined regions: red diamonds for region A,
blue squares that is for region B, and green crosses for region C. The coloured solid 
lines represent the predictions made by the models in the MB thin-shell approach, 
while the black dashed line represents the prediction of the model in the 
RB thick-shell approach (see Section 3 for a explanation of the models).}
\label{LHb}
\end{figure}

\subsection{Description of the measured radial profiles}

The co-addition of the emission from the fibres included in each annulus 
allowed us to find the radial variation of different ratios based on
the reddening-corrected emission-line intensities.
The different spatial behaviour observed in these ratios for different angular
positions has motivated a separate analysis.

The radial profile of the \hb\ surface
brightness relative to the \hb\ emission peak, as plotted in Fig. \ref{LHb}, 
has its maximum in the central position, and it decreases
with distance in the three regions, although in a more irregular manner in region C.

\begin{figure*}
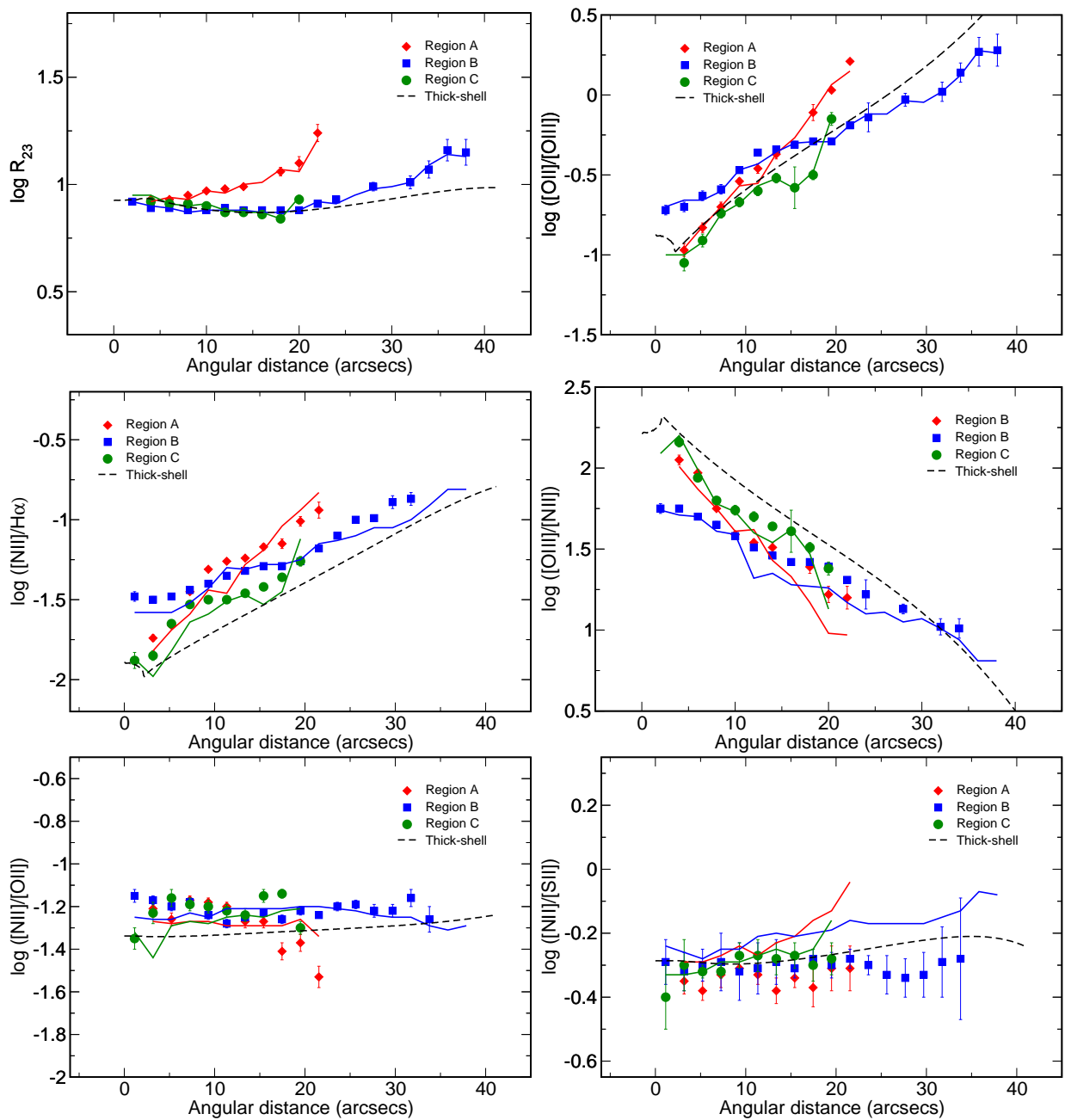


\begin{minipage}{180mm}
\centerline{
\psfig{figure=R23.eps,width=8cm,clip=}
\psfig{figure=o2o3.eps,width=8cm,clip=}}
\centerline{
\psfig{figure=n2.eps,width=8cm,clip=}
\psfig{figure=O3N2.eps,width=8cm,clip=}}
\centerline{
\psfig{figure=n2o2.eps,width=8cm,clip=}
\psfig{figure=n2s2.eps,width=8cm,clip=}}

\caption{Radial profiles of several emission-line ratios based
on the optical reddening-corrected emission lines across the annuli in the three defined regions. 
>From up to down and left to right: \rdostres, [\oii]/[\oiii], [\nii]/\ha\, [\oiii]/[\nii],
[\nii]/[\oii], and [\nii]/[\sii]. Symbols and lines are the same, as defined in Fig.\ref{LHb}.}
\label{opt_prof}

\end{minipage}
\end{figure*}

In Fig. \ref{opt_prof}, the radial variation of some ratios involving
the measured optical emission-lines are shown.  The
\rdostres\ parameter is defined as the sum of [\oii] $\lambda$ 3727
\AA\ and [\oiii] $\lambda$$\lambda$ 4959,5007 \AA\AA\ relative to
\hb\ intensity \citep{pagel79} and is used as an estimator of
metallicity, despite its bi-valuated behaviour. Its radial profile,
shown in the upper left panel of Fig. \ref{opt_prof}, is quite
different in the three regions: it is uniform around log
\rdostres\ $\approx$ 0.9 in the inner annuli of region B and in the
whole region C, but it increases with distance in region A and the outer
annuli of region B, reaching values larger than log
\rdostres\ $\approx$ 1.3.

The ratio of oxygen emission lines [\oii]/[\oiii], shown in
upper right panel of Fig.\ref{opt_prof},
depends mainly on the ionisation parameter (\citealt{diaz}, $\log U$; i.e.,
the ratio between ionising photons and  particle density, in the
case of matter-bounded nebulae also depends strongly on the thickness of
the gas envelop.
The radial profiles have similar
behaviours in the three regions with the same average value
being lower at closer distances to
the ionising source (e.g., with larger $\log U$) but having a larger slope 
in regions A and C and lower in region B.

The emission-line ratio [\nii] $\lambda$ 6584\AA/\ha, defined as N2,
is used as an estimator of metallicity (e.g., \citealt{storchi}),
although it also depends on the ionisation parameter \citep{pmd05} and on
the nitrogen abundance \citep{pmc09}. As can be seen in the left middle
panel of Fig. \ref{opt_prof}, [\nii] $\lambda$ 6584\AA/\ha  increases
radially from the inner to the outer annuli in all three regions, although the average value
is lower in region C.  The opposite behaviour is seen for the ratio of
[\oiii] at $\lambda$ 5007 \AA\ and [\nii] at $\lambda$ 6584 \AA,
(middle right panel of Fig.\ref{opt_prof}), known as O3N2, and is also
used to estimate metallicities \citep{alloin}.  In region B, it is
found for both the lower average value and variation.

In the case of ratios that depend only on low-excitation emission lines, the observations in MI11 were  confirmed. Both
[\nii]/[\oii] and [\nii]/[\sii], which are also known as N2O2 and N2S2, respectively, and 
which are proposed as estimators of
N/O \citep{pmc09}, remain constant across all the three regions,
as shown in lower panels of Fig.\ref{opt_prof}.

\begin{figure}
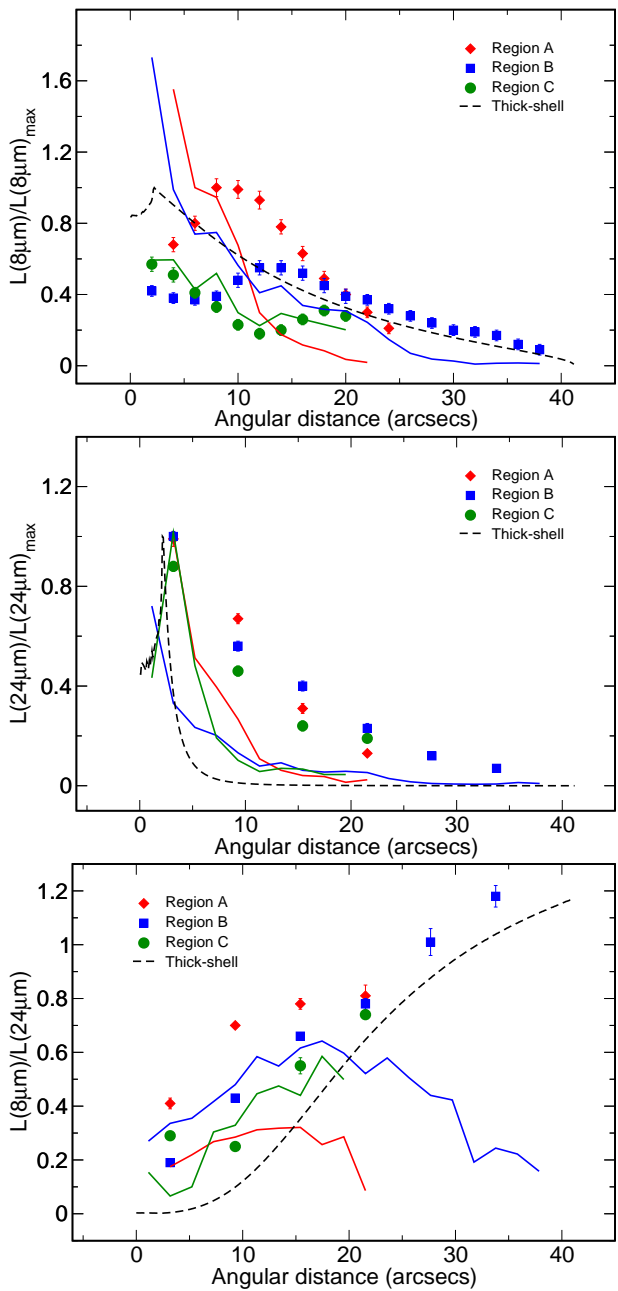

\psfig{figure=8mu.eps,width=8cm,clip=}
\psfig{figure=24mu.eps,width=8cm,clip=}
\psfig{figure=8to24.eps,width=8cm,clip=}
\caption{Radial profiles of the relative surface brightness
of the {\em Spitzer} emission bands at 8~\mi\ 
and 24~\mi\ and the corresponding ratio
across the annuli in the three regions with symbols
as defined in Fig. \ref{LHb}.}
\label{IR}
\end{figure}

Regarding the observed surface brightness of the
{\em Spitzer} bands at 8~\mi\ and 24~\mic\ , as shown in Fig.\ref{IR}, both present
maxima at the central annuli of the three regions with decreasing
radial profiles, although  the maximum 8\mic\ emission 
is slightly displaced towards outer position in regions A and B.
The ratio of surface brightness 8~\mic/24~\mic\, sampled at the
same spatial resolution increases radially
in the three regions, from values around 0.2 in the central position
to values higher than 1 for the further ones, which agrees with the
average values of this ratio found by \cite{bendo} in a sample of 
\hii\ regions in spiral-discs. The average value of this ratio, however,
is sensibly lower in region C.

\subsection{Dust mass and extinction in NGC~588}
\label{dustIR}

Mid- and far-IR observations allow us to estimate the total amount of dust within 
the G\hii R
and obtain an estimate of the extinction.
The dust mass derived from these observations can be used to derive a 
dust-to-gas mass ratio that can be later compared with results from the models. 

We assumed that most dust mass is in the form of grains in thermal equilibrium, 
thus the dust mass can be computed for a given flux ($F_\nu$) using the following expression: 

\begin{equation}
F_{\nu}(T) = \frac{\rm M_{dust}}{4\pi D^2} \cdot 4\pi  \kappa_{\nu}B_{\nu}(T)
\label{eqdust}
\end{equation}
where $F_{\nu}(T)$ is the observed flux, $B_{\nu}(T)$ is the Planck function, 
$\kappa_{\nu}$ is the mass absorption coefficient ($\kappa_{160\mu m}$=24.6 
cm$^2$ $\cdot$ g$^{-1}$ \citep{Lisenfeld} with  $\kappa_{\nu}\sim\nu^{\beta}$, 
$\beta=2$, which is typical of interstellar grains \citep{1984ApJ...285...89D}, and $D$ is 
the assumed distance of M33 (840 kpc), \citealt{Freedman}). To derive the dust mass,
we need to know the dust temperature, which can be estimated from the $F$(70~\mi)/$F$(160~\mi) 
ratio (e.g., \citealt{Tabatabaei:2007p664}). 
We estimated the dust temperature by using the 70~\mi\ and 160~\mi\ images of M33 from 
{\em Spitzer} \citep{Verley:2007p574}, whose emission was integrated for NGC~588. 
Assuming  an upper limit, where the entire 70~\mi\ emission comes from big grains, 
this temperature is $\approx$ 29 K, and assuming that only 60\% of the 70~\mi\ 
emission comes from big grains, we estimated a temperature of $\approx$25\,K. 
The dust temperature estimate for NGC~588 is slightly higher than the one derived 
in \cite{xilouris}, which uses two modified blackbodies that include the emission from 24\mi\ to 500\mi.

By taking values for the dust temperature of T=25, 27, and 30\,K, we estimated the 
dust mass using the emission at 160~\mi\ in Eq.\,\ref{eqdust}, and we found 
$\rm M_{dust}\approx 500\rm\,M_{\odot}$, $\rm M_{dust}\approx380\rm\,M_{\odot}$, 
and $\rm M_{dust}\approx 270\rm\,M_{\odot}$, respectively. 
The total H{\sc i} mass corresponding to the \hii\ region was derived from the 
H{\sc i} map given in \cite{2010A&A...522A...3G}. We integrated the H{\sc i} 
emission by using the
same 70\arcsec diameter aperture to obtain the 70~\mi\ and 160~\mi\ fluxes
and derived a total H{\sc i} mass of 6.5$\times 10^{4}\rm\,M_{\odot}$. 
Since the mass of the molecular cloud associated with NGC~588 is negligible 
(\citealt{2003A&A...398..983B} and \citealt{Verley:2007p574}) we can consider 
the M(H{\sc i}) to be representative of the total gas mass. With this 
approximation, we derived values for the dust-to-gas ratio in NGC~588 
of $\sim7.7\times10^{-3}$, $\sim 5.9\times10^{-3}$, and $\sim 4.1\times10^{-3}$ 
for the three dust mass estimates, respectively (500 M$_{\odot}$, 380 M$_{\odot}$, 
and 270 M$_{\odot}$). These estimates for the dust-to-gas ratio are close to 
the Galactic standard value ($\sim6.7\times10^{-3}$, \citealt{1984ApJ...285...89D}). 

We can also derive the inner extinction, which is the amount of dust
produces in the \hii\ region.  Using the dust temperature estimated
above and the integrated flux at 160\,\mi\ for the \hii\ region, we
can predict the dust opacity at 160\,\mi, $\tau_{160}$$_{\mu}$$_m$ 
by following the equation:

\begin{equation}
\tau_{160}=F_{160}(T)/\Omega\,B_{160}(T), 
\label{eqext}
\end{equation}
\noindent where $\Omega$ is the solid angle covered by the region. 
Assuming $\tau_{H{\alpha}}$ $\simeq\,2200\times\tau_{160}$$_{\mu}$$_m$
\citep{Tabatabaei:2007p664} from standard dust models for the diffuse emission, 
we obtain an estimate of the extinction at \ha. 
We derived values for $\tau_{H\alpha}$ of 0.06, 0.04, and 0.03, which is equivalent to extinctions (A(\ha)=1.086$\times\tau_{H{\alpha}}$) of 0.06, 0.05, and 0.03 for dust temperatures of 25K, 27K, and 30K, respectively.   
This range for the extinction agrees within the uncertainties of the intrinsic extinction that is derived using
the H$\alpha$/24~\mi\ ratio for NGC~588 \citep{relano09}. However, it must be
noted that this value is much lower than the average extinction value derived
from the Balmer decrement both from IFS (MI11) and
other optical integrated observations (e.g. \citealt{melnick79}, \citealt{vg86},
\citealt{melnick87}) and which point to values of the inner extinction 
in the range A(\ha) $\approx$ = 0.3 - 0.7. This disagreement could be consequence
of different extinction properties for the gas in the \hii\ region and the dust
emitting in the IR.

\begin{figure*}
\begin{minipage}{170mm}
\centerline{
\psfig{figure=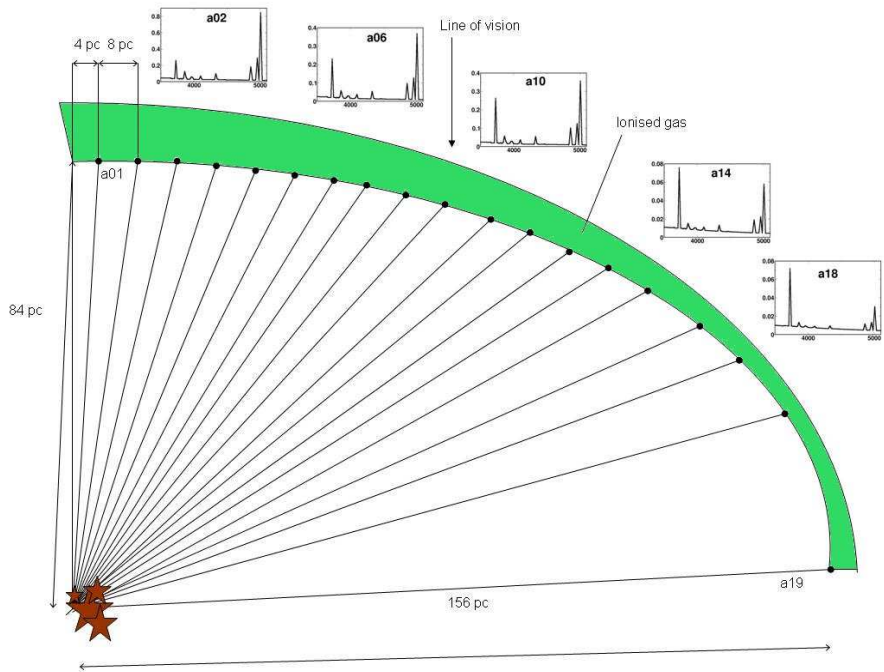,width=12cm,clip=}}
\label{mod_scheme}
\caption{Sketch is not to scale for the adopted geometry in  MB thin-shell approach 
in the region B (W half) of NGC~588. The line of vision indicates the orientation of the detectors used to acquire 
the data. Black points represent the positions of the modelled annuli plotted in Fig.~\ref{annuli}.
We also show  the model-resulted optical spectrum for some annuli to illustrate the
radial variation of [\oii], [\oiii], and \hb\ across the nebula. Notice that
 in Fig. 1, photons only cross once in the shell although annuli look concentric.
}
\label{sketch}
\end{minipage}
\end{figure*} 

\section{Model description}
\label{models}
Photoionisation models were performed to reproduce the observed relative fluxes 
of the brightest optical emission-lines and {\em Spitzer} mid-IR broad filters that were centred 
at 8 \mic\ and 24~\mic\ in all annuli of the three defined regions. 
Improving the models described in \cite{mod595} for NGC~595,
where the observed [\oii]/\hb\ and [\oiii]/\hb\ emission-line ratios were used,
we also compared the models with the nebular ratios of [\nii] $\lambda$ 6584 \AA/\ha\ 
and [\sii] $\lambda$$\lambda$ 6717,6731 \AA\AA/\ha. 

To do so, two different extreme approaches were envisaged: i) the
observed properties in each annulus were considered to be independent of
the properties of the rest of the nebula, so a different model was
computed for each one of these annuli. This methodology is useful for
seeking azimuthal variations in the properties of the gas and the
dust.  The resulting geometry of the models under this assumption is a
matter-bounded (MB) thin shell. ii) The observed properties of the gas
are a consequence of the projection of an emitting gas-sphere onto the
sky. This approach is useful for studing possible projection effects.  In
this case, a single model was done to reproduce the observations and a
radiation-bounded (RB) thick shell is obtained in this case.

In both approaches, models were made using the code {\sc Cloudy} v.10.00 \citep{ferland},
which models the effects of a radiation source on a one-dimensional gas and dust distribution.
Input gas-phase abundances were used 
to match the derived metallicity  from optical observations
and assumed chemical homogeneity, as suggested
in MI11.
In the thin-shell approach the observations were reproduced using the 
value derived by \cite{jamet05}, which includes 
He, O, and S elemental abundances that are derived from optical, collisionally excited lines
and the measurement of the corresponding electron temperatures.  In the thick-shell
model, a slightly higher oxygen abundance was required [12+log(O/H) = 8.3]
to reproduce the observations.
In both approaches, a same value of N abundance was used to reproduce the radial
profiles involving [\nii] emission lines [12+log(N/H) = 6.9]. 
The abundances for the rest of elements were rescaled by taking  the difference to the oxygen abundance in the solar photosphere as 
a reference as given by \cite{asplund}.

Regarding the ionising spectral energy distribution (SED)
used in both approaches, we adopted the conclusions of the 
spectrophotometric star-by-star study of this G\hii\ R made by 
\cite{jamet04}. These authors conclude that the SED 
of the three observed ionising clusters is dominated
by the most massive stars in the southern cluster, which hosts 
two WN stars. Later, \cite{jamet04} adopted an effective
temperature of 41\,000 K as equivalent for the three clusters.
Therefore, we used  the SED of a 
single WM-Basic O star \citep{wm} with metallicity Z = 0.5$\cdot$Z$_{\odot}$ for our models (the closest 
available to the value that corresponds to the measured oxygen gas-phase abundance), 
log(g) =  4.0, and effective temperature equal to 41\,000 K. 
We checked  the effect of a Starburst99
\citep{leitherer} SED in our models that corresponds to the equivalent cluster
age proposed in \cite{jamet04} (around 4.3 Myr), 
which is attained by means of, among other techniques, colour-magnitude diagrams. 
However, we were not able to  satisfactorily reproduce the radial profile of sny of the
observed properties, whose spatial distribution we tried to fit. This is owing
to the much lower relative number of ionising photons in a SED for
this age in relation with the two dominant massive stars.
This IMF sampling effect in this low massive G{\hii}R is well described in \cite{cervino03}
and \cite{jamet04}. In any of these cases, although all the observed emission-line profiles
are well fit using the assumed single star SED, some effects from the low
massive stars to the low-excitation emission-lines are not be negligible.

\begin{table}
\label{table2}
\begin{minipage}{80mm}
\caption{List of observable and physical properties needed for the
models under the MB thin-shell and 
RB thick-shell approximations. During the
iterative process, they are considered as free (no restriction in each
iteration; F), constrained (the model searches a convergence to the expected
values; C), and input (fixed values from the beginning; I).}
\begin{tabular}{lcc}
\hline
\hline
 & MB Thin-shell & RB Thick-shell \\
\hline
L(\hb) & C & C  \\
I([\oii])/I(\hb) & C & C \\
I( [\oiii])/I(\hb) & C & C \\
I([\nii])/I/\ha) & C & C \\
I([\sii])/I/\ha) & C & C \\
L(8\mic) & F & F \\
L(24\mic) & F & F \\
gas density & I & F \\
Q(H) & F & I \\
dust-to-gas ratio & F & I \\
abundances & I & I \\
SED & I & I \\
inner radius & I\footnote{free in the first annulus} & I \\
outer radius & F & F \\
filling factor & F & I \\
\hline

\hline
\end{tabular}
\end{minipage}
\end{table}

In Table 1, we list the different parameters
considered in the two geometrical approache, by distinguishing
between the following input, if they are fixed in the models; free, if they vary
to find the better solutions; and constrained, if they are
fit to the observed values by means of varying the free
parameters. Details are explained in the following sub-sections.

\subsection{MB thin-shell geometry models}

We used the same methodology and geometrical approach assumed for the
modelling of NGC~595 that was carried out by \cite{mod595}.
This consists of the direct comparison between the transmitted spectra predicted by the
models in each annulus, as defined over the shell, and the  
IFS observed spectra in the same regions.
As {\sc Cloudy} models  result in a closed geometry around the ionising sourceby default  and
observations are analysed in a plane, all resulting magnitudes were
divided by the corresponding area before their comparison (i.e., 
the surface of a sphere in the case of models and the corresponding angular size
of the annuli for the observations).
This procedure gives a qualitative rather than a quantitative analysis, 
but it constitutes a solid approach to study the azimuthal variations of the physical
properties of gas and dust as a function of the distance to the ionising source in each 
part of the nebula.

A model for each annulus was calculated assuming that a complete
ionisation structure is formed for different positions of the shell.  
A set of initial input model conditions was imposed in each model, and then an automatic
iterative process was carried out by the code to fit the observed features, including the
emission-line intensities relative to \hb\, of [\oii] at $\lambda$ 3727 {\AA}, [\oiii] at $\lambda$ 4959, 5007 {\AA},
[\nii] at $\lambda$ 6584 \AA, and [\sii] at $\lambda$ 6717, 6731 \AA, 
and the \hb\, surface luminosity relative to the maximum luminosity 
measured in the \hii\, region.

Since a hollow shell is considered, the distance between the inner
face of the shell and the ionising source in the central annulus
cannot be taken in the models simply as the projected angular
distance, which is 4 pc (corresponding to 1 \arcsecond\ at the
assumed M33 distance).  Instead, a larger distance for the inner radius
in this closest annulus is required to optimise the fitting
of the observed quantities in the models. In this case, the best
agreement between the observed and modelled emission-line ratios was
found by assuming a distance of 84 pc between the inner face of the gas
and the ionising source. The difference between the distance found by
the models and the observed projected distance in this annulus can be
interpreted as an unseen component of the distance in the direction
perpendicular to the plane of projection. However, the inner radius
for this annulus is a free parameter of the model and must be
considered as an arbitrary value found to reproduce the observations
under this geometrical assumption.  In contrast to the innermost annuli,
in the outermost this deprojection does not require models and observation do not agree.  Therefore, the considered
factor of deprojection for the furthermost annulus is 1 (i.e. the
distance measured in the projected image and the inner radius of its
corresponding model is the same, which is 156 pc in region B).  In
this way, the ratio between the adopted and projected distances is not
uniform in the intermediate annuli, but it gradually decreases
by assuming the geometry of a hollow ellipsoid of revolution.  In
Fig. \ref{sketch}, this can be seen by the resulting shell with their different
modelled positions, which correspond to the annuli shown in Fig. 1 in a
cut perpendicular to the plane of the projected sky for region B.
However, it must be kept in mind that the ratio between semi-axes in
this resulting geometry is arbitrary, as no reliable information about
the deprojection factor or the inner radius can be obtained using this
method.
A constant density of 50 particles per cm$^3$ was assumed, 
according to the values measured using the emission line ratio of [\sii]
6717\AA/6731\AA, 
which ranges between 10 and 100 particles per cm$^{3}$ in the considered regions
(MI11).

We left the following as free parameters: the number of ionising photons, the filling
factor, the thickness of the gas shell, and the amount of dust, which
is varied in each iteration of the model as a means to find the best
agreement with the observed \hb\ surface brightness and emission-line
ratios. A different number of ionising photons was considered in each
model to test the robustness (i.e., non-degeneracy) of the
solution in each model.
The dust-to-gas ratio must also be considered to reproduce the observed emission
because the dust heating affects thermal balance of the gas. It was also left
as a free parameter to study possible spatial variations across the field of view.
We adopted the default grain properties of {\sc Cloudy} v10.00, which essentially has  the 
properties of the ISM and follows a MRN \citep{mrn} grain size distribution. 

In all models, the resulting filling factor and thickness lead to a 
plane-parallel MB geometry, where a large fraction of the ionising 
photons emitted by the central cluster escape to the outer ISM of the galaxy. 
The average $\chi^2$ in the four optical emission-line ratios
(i.e., [\oii]/\hb, [\oiii]/\hb, [\nii]/\ha, and [\sii]/\ha) and the relative \hb\ flux
is better than 0.2 in all cases. The results of these models of the three
studied regions for relative \hb\ surface brightness, optical emission-line ratios,
and mid-IR emission are plotted as solid lines in Figs. 2-4.

\subsection{RB thick-shell geometry model}

\begin{figure}
\psfig{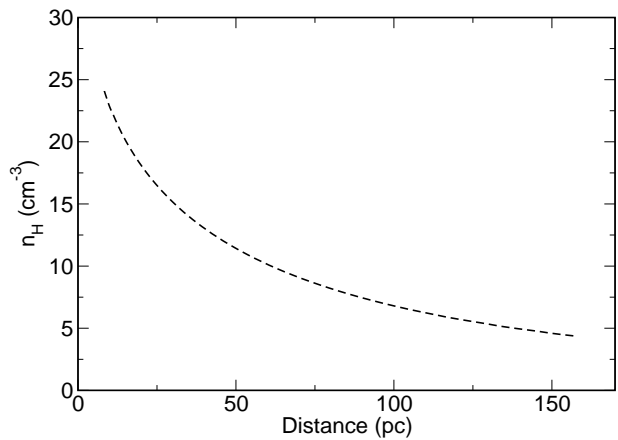}
\caption{Particle density variation across the radius of the nebula considered by 
the model in the thick-shell approach.}
\label{nH}
\end{figure}

As an alternative to the models presented in the previous subsection,
we explored the possibility of a spherical distribution of the gas. A
sphere around the central source filled with gas that assumes a certain
density law naturally shows a decrease in the surface brightness
due to the diminution of the column with the angular distance to the
centre of the object. Nevertheless this radial variation would  not
exactly follow the observed shape of the surface brightness decrease,
so we need to also change the density according to the radius. We used
the pseudo 3D code {\sc pyCloudy}\footnote{Available at
  https://sites.google.com/site/pycloudy/} (the python version of {\sc
  Cloudy 3D}, \citealt{C3D}) to compute a spherical model and to
determine the values of the emission line intensities when projected
on the plane of the sky.  The {\sc pyCloudy} code generates {\sc
  Cloudy} input files corresponding to different angular directions
(from the point of view of the ionising source) and runs the
corresponding 1D model, which is later integrated in the line of
vision for different impact parameters to get the observed radial
variation.
We determined the density law $n_{\rm H}(r)$ by trial and error, so
that the global variation of the surface brightness is reproduced. The
resulting density law is shown in Fig. \ref{nH} and has the following
form,
\begin{equation}
n_H(r) = 40\cdot \exp(-(r/32.4)^{1/2})
\end{equation}
with $n_H$  units of cm$^{-3}$ and $r$ in units of pc. The inner radius 
of this model was arbitrarily set to 8.33 pc.

As in the previous approach, the thick-shell model has a unique
central photoionising source that dominates the equivalent SED in this
G\hii R.  The number of ionising photons in this model, Q(H) =
10$^{50.73}$ s$^{-1}$, was set to agree with the value derived from
the extinction-corrected total H$\alpha$ luminosity derived by
\cite{relano09} [log(L(H$\alpha$) = 38.92 $\pm$ 0.09 erg/s].  Dust is
included in the model by using the same standard properties described for
the thin-shell model, and no attempt was made to change the dust
content with the radius. A constant filling factor of 0.055 was used
to match the observed size and density of the region. The model is
RB. The emergent spectrum is calculated when the electron temperature
is lower than 4\,000 K, which happens for an outer radius of 160
pc. The decrease of the ionisation relative to the radial distance to
the centre of the object is naturally reproduced by the decrease of
the ionisation parameter due to the dilution of the radiation when
moving away from the central source. The aim of this model is only to
reproduce the global variations of the observables with the radial
distance; its simpleness can obviously not reproduce the differences
observed in the regions A, B, or C.  These differences are thus thought
to be associated with local variations in the assumed radial laws for
the density and filling factor. The results of this model in regards to
\hb\ relative surface brightness, optical emission-line ratios and
mid-IR emission is shown as a dashed line in Figs. 2-4.

\section{Results and discussion}

\subsection{Geometrical implications}


\begin{figure}
\psfig{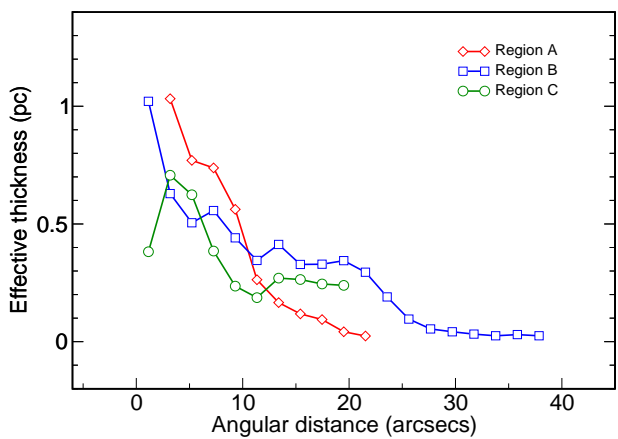}
\caption{Radial profile of the effective thickness in NGC~588 predicted 
by the MB thin-shell models in the three defined regions with symbols as
defined in Fig. \ref{LHb}.}
\label{Reff}
\end{figure} 

\begin{figure}
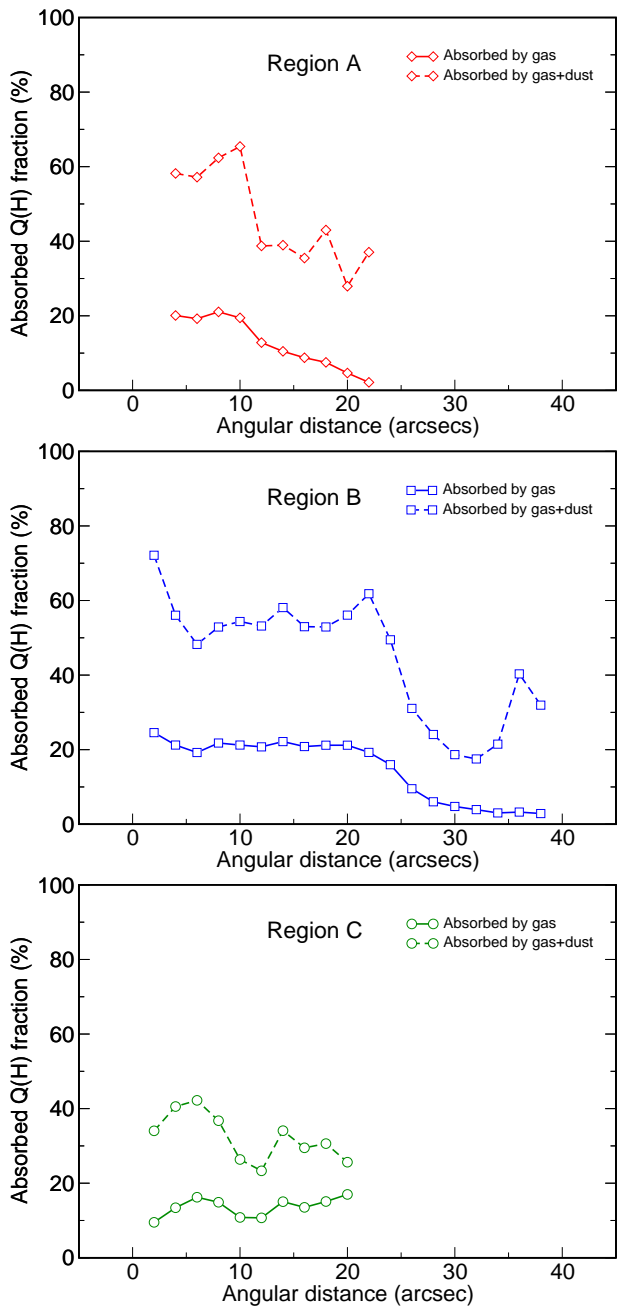

\psfig{figure=escape_A.eps,width=8cm,clip=}
\psfig{figure=escape_B.eps,width=8cm,clip=}
\psfig{figure=escape_C.eps,width=8cm,clip=}
\caption{Radial profile of the fraction of ionising photons 
absorbed by the gas (solid line) and
by both the gas and the dust (dashed line), as predicted 
by the MB thin-shell models for
regions A, B, and C from up to down, respectively.}
\label{escape}
\end{figure} 

\begin{figure}
\psfig{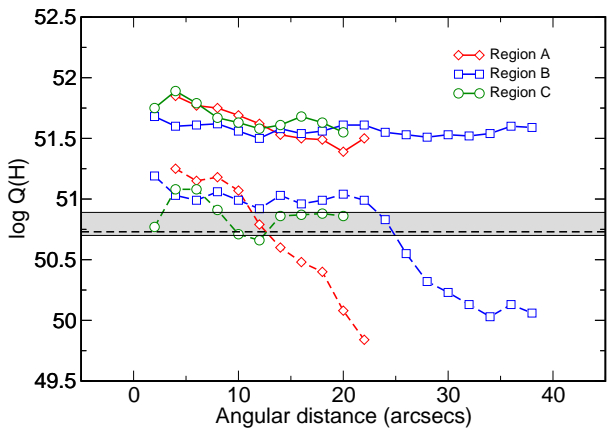}
\caption{Number of ionising photons required by the MB thin-shell models emitted by
the ionising source for each region (solid line) and absorbed by the gas (dashed line).
The black dashed line represents the input number of ionising photons of the RB thick-shell
model, while the grey band is the number derived from the total H$\alpha$ 
extinction-corrected luminosity measured by Rela\~no \& Kennicutt (2009). }
\label{QH}
\end{figure}

Although the models under the two different assumptions described above 
are able to reproduce most of the 2D observed structure of NGC~588, the
final geometries considered in them are very different.
On one hand, a thin-shell geometry for the gas surrounding the ionising star clusters and
following the shape of an ellipsoid in revolution, as sketched in
a vertical cut in Fig. \ref{sketch}, turns out 
fundamental to correctly fit the distribution for most of
the observed optical properties in this G\hii R.
On the other hand, a thick-shell geometry in a sphere appears naturally
when a single 3D model projected in two dimensions is considered instead.

The relative distance between the
three main stellar clusters in NGC~588 reported by
\cite{jamet05} results in a much lower distance
between them and the outer radius of the gas-shell, 
thus, minimising the effects of the spatial distribution 
of the stars. According to \cite{ercolano}, this distribution can have
non-negligible effects on the ionisation and thermal  structure of the gas.
Although  the SED is dominated by a few massive
stars located close to each other in this nebula, the distribution of the three observed clusters
can be relevant for the properties of the that ate gas closer to the stars in the
RB thick shell approach, but no direct isolated observations
of the gas in these positions is available for such an analysis.


While  we fixed the filling factor 
to match the observed size of NGC~588 in the thick-shell approach,
all the models for the different positions of 
the thin-shell were calculated by using  the filling factor and the outer radius of the shell as free parameters,
among others, in
each one of the modelled annuli. 
The resulting models present a degeneracy in
these two parameters to study them consistently, and
to establish comparisons between the different thickness in each annulus,
it is necessary to define the {\em effective thickness} as
\begin{equation}
\Delta R_{eff} = \epsilon \times \frac{V(r_i,r_o)}{S(r_o)},
\end{equation} 
\noindent where $\epsilon$ is the filling factor, $V$ is
the volume of the shell, which depends on both the inner and the outer
radii, $r_i$ and $r_o$, and $S$ is the outer surface of the shell. 
In Fig. \ref{Reff},  the resulting
effective thickness is shown for the three  defined regions above as
a function of the distance between the ionising source
and the inner face of the shell. The effective thickness is larger
in the inner annuli of regions A and B, and it then decreases for the outer annuli,
while  the thickness is more irregular in region C. 
This thickness variation is required by the models to explain the
observed shape of the \hb\ surface brightness.

As a consequence of the resulting MB geometry for this solution in all
annuli, a fraction of hydrogen ionising photons, $Q(H)$, escape from
the gas envelope.  In Fig. \ref{escape}, the predicted fraction of
absorbed photons from only gas (in solid line) and from both the gas
and the dust (in dashed line) for each one of the three regions as a
function of the distance to the ionising source is shown.  As can be
seen, according to these models, the fraction of photons absorbed by
the gas is larger in the inner annuli of regions A and B, where the
effective thickness is larger, with values around 25\% and down to
only 4\% in the outer annuli of these regions. In contrast, this
fraction is keept around 20\% in region C but with great fluctuations
due to the variations in the effective thickness. Taking the fraction
of the photons absorbed by the dust into account, we can calculate the
fraction of photons escaping from the nebula.  The fraction of these
escaping photons changes from around 20\% in the inner annuli of
regions A and B and up to 70\% in the outer annuli. For region C, this
fraction oscillates between 30\% and 60\%.  Using a surface-weighted
mean, we estimate the mean fraction of hydrogen ionising photons
absorbed by the gas (10 \%) and those escaping from the nebula (60\%).
Taking these factors into account, the number of ionising photons
absorbed by the gas in these models is log Q(H) = 50.70 which agrees
with the total H$\alpha$ luminosity measured by \cite{relano09} for
this G{\hii}R.  The total number of ionising photons and the fraction
of them that is absorbed by the gas is plotted for each model under
this assumption in Fig. \ref{QH}.  In contrast, the number of ionising
photons in the thick-shell approach is totally absorbed by gas and
dust, and no photon-leakage is predicted, therefore agreeing with the
measured H$\alpha$ luminosity in this G{\hii}R.

\subsection{Functional parameters}

According to the more classical vision in the study of
{\hii} regions, there are three so-called functional parameters,
which control the ionisation and thermal inner structure
of the gas (e.g., \citealt{mcgaugh}, \citealt{pmd05}): the metallicity ($Z$),
the ionisation parameter ($\log U$), and the
equivalent effective temperature ($T_*$).
As shown above, the relative geometry among the gas,
the dust, the stars, and other important parameters, such
as the dust-to-gas ratio, also have very important
roles that become evident with the study of the 
spatial distribution of the observed properties.
Nevertheless, the study of the three classical
functional parameters is still fundamental for
the global understanding of the physical processes
governing the interplay between gas, dust and radiation
in ionised gaseous nebulae. 

Although in
MI11, for instance, a guess about the behaviour of some
of these parameters, as derived from the variation of the relative
intensity of the brightest optical emission-lines in NGC~588 is proved ; all the used 
strong-line methods are based on calibrations of integrated observations or models,
and a poor knowledge about their behaviour in spatially resolved
regions is attained. Therefore, a thorough check of the
main strong-line methods in this context by means of tailor-made models 
is compulsory to confirm the conclusions obtained in MI11 and to
explore the validity of these methods for spatially-resolved G\hii\ Rs.

\subsubsection{Metal content of the gas}

The amount of chemical species that are heavier than helium in
the gas-phase is one of the factors that most
affects the cooling rate of the gas and thus 
the emissivity of the optical emission-lines.
At present, in the case of NGC~588, there is no observational evidence of any chemical inhomogeneity.
Our models support this result since the radial profiles of 
all optical emission-lines obtained using
IFU PMAS are reproduced under two different adopted geometries.
The uniform oxygen abundance considered in the models
[12+log(O/H) = 8.16 (thin-shell), 8.30 (thick-shell)] agrees with the values derived
by different authors (\citealt{vilchez88}, \citealt{jamet05}) by means of
collisional emission-lines and an estimate of the electron temperature from
auroral-to-nebular appropriate emission-line ratios.
Nevertheless, it has been found by \cite{jamet05} that a non-negligible disagreement
between the [\oiii] electron temperature
derived from the integrated ratio of [\oiii] $\lambda$ 5007 \AA\
and $\lambda$ 4363 \AA\ in the optical and the ratio between
$\lambda$ 5007 \AA\ and the IR emission-line at $\lambda$ 88~\mi, which has been 
taken from {\em ISO} observations, is about 3\,000 K lower.
However, despite that a realistic SED was not used, 
we do not find  any difference between the electron temperature derived
from [\oiii] optical and mid-IR emission lines in our models.
This result agrees with those from one-dimensional
tailor-made models for the same \hii\ region made by \cite{jamet05} who varies
their assumptions on the geometry, density, and adopted SED.

Strong-line methods used to derive $Z$ (e.g., \rdostres, O3N2,
and N2), all present variations across the nebula that can be
explained without resorting to chemical inhomogeneities. According to
\cite{pmd05}, most of them have non-negligible dependences on the
other functional parameters.  As shown in the
upper left panel of Fig. \ref{opt_prof}, the case of \rdostres\ remains nearly uniform in
the inner annuli of region B, and all region C.  As was
already pointed out by MI11, \rdostres\ in this G\hii\ R is at the
value of the so-called {\em turnover} region (i.e., when
\rdostres\ has a maximum value, and both upper- and lower-branch
calibrations join (e.g., \citealt{mcgaugh})).  This means that this
parameter cannot be used to find out metallicities within the range
8.0 - 8.4. In addition,  the
\rdostres\ parameter reaches values even higher than the maximum
obtained for different sequences of photoionisation models and
compilations of integrated observations in the outer annuli of regions A and B.  According to some authors
(e.g., \citealt{dopita}), the very high fluxes of [\oii] and [\oiii]
relative to \hb\ can be caused by the presence of high-velocity
gas shocks.  To study the possible presence of shocks, we looked for
them by two means: i) we searched for emission of the good shock
tracer [\oi] 6300 \AA, which is indicative of shocks
(e.g. \citealt{dopita}). At the distance of M33 and at the spectral
resolution of the IFU data, it was not possible to properly deblend the
emission of the [\oi] 6300\ \AA\ sky line from a putative emission
from NGC~588.  Therefore, this search relies on a good background
subtraction. The only part of our field of view where this could be
done was in the lower left tile, since it  simultaneously samples the
edge of the G\hii R and the emission from the rest of M33. We coadded
the emission that corresponds to several spaxels of the region and to the
background in this tile and subtracted them. The search for the line
was unsuccessful: ii) we looked for spaxels with line widths typical
for those found in shocks ($\gsim$ 100~km~s$^{-1}$). No spaxels with
line widths larger than the instrumental one
(i.e., $\sigma_{instr}\sim66$~km~s$^{-1}$) were detected. Therefore, no
observational evidences support an important role of shocks in this
region.  The very high values of \rdostres\ could only be reproduced
in the thin-shell approach without considering either chemical
inhomogeneities or high-velocity gas-shocks and by only assuming changes
in the geometry and the dust-to-gas ratio in the observed regions.


\begin{figure}
\psfig{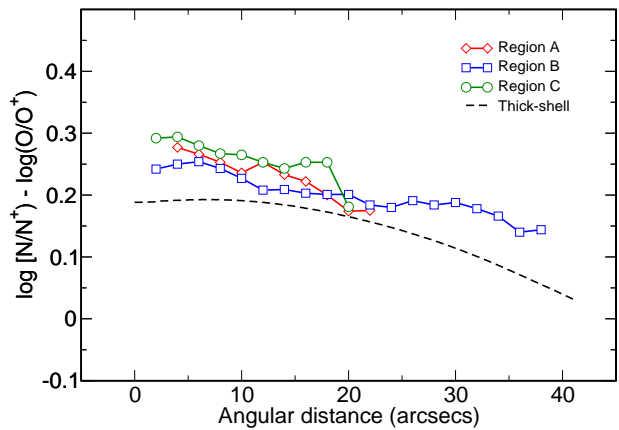}
\caption{Radial variation of the difference between log(N/N$^+$), the ICF for N$^+$, and
the usually adopted ICF, log(O/O$^+$), as predicted by the 
thin-shell models in the three studied
regions and by the thick-shell model. 
All ionic abundances were taken from the models as the integrated 
density-weighted values over the radius.}
\label{icfN}
\end{figure} 

Other strong-line methods to derive $Z$ are based on
[\nii] emission lines, as it is the
case of N2 and O3N2. The radial profiles of these two parameters
can be seen in left and right middle panels of Fig. \ref{opt_prof},
respectively. They both present a large
variation from the inner to the outer annuli in the three regions.
According to \cite{pmc09}, all strong-line estimators of the
metallicity based on [\nii] emission lines have a non-negligible dependence of the
nitrogen-to-oxygen ratio (N/O). However, as in the
case of $Z$, all our models correctly reproduce the inner variations
of all quantities involving [\nii] emission-lines by
assuming a uniform value for N/O.
Therefore, the  variations of N2 and O3N2, can only be explained by
invoking geometrical effects, which also lead to variations in the ionisation parameter.
Indeed, the two strong-line methods used to
derive N/O, N2O2 and N2S2, which do not depend on log $U$), 
do not present large variations, as can be seen in the left and right lower 
panels of Fig. \ref{opt_prof}, respectively.

The N/O value derived from thin-shell models is
0.15 dex larger than the value reported by \cite{jamet05}
[log(N/O) = -1.39 $\pm$ 0.09]. In contrast, this value agrees 
with the value obtained from the thick-shell model.
The cause of this 
disagreement  for the N/O abundance ratio can be sought in the
corresponding ionisation correction factor (ICF) for nitrogen. The most 
widely used expression for this ICF is based on the assumption:
N/O $\approx$ N$^+$/O$^+$, which leads to
\begin{equation}
\textrm{ICF}(N^+) = \frac{N}{N^ +} \approx \frac{O}{O^+}.
\end{equation}
\noindent This expression is supported by sequences of photoionisation
RB models at different N/O ratios by \cite{pmc09}. Nevertheless, as can
be seen in Fig. \ref{icfN}, there is an offset of $\sim$ 0.2 dex on 
average for all the annuli of the three studied regions in NGC~588,
which gives place to the larger total N/O ratio  in the thin-shell model
relative to the classical approximation. This can be due to the cut 
in the ionisation structure 
in several MB gas configurations and, thus, vary the
usual expressions for ICF that are considered in models of a less complex geometry.
In the thick-shell model, N/N$^+$ is 0.1 dex
higher than O/O$^+$ on average.

In both models, the derived total nitrogen abundance
[12+log(N/H) = 6.9] is consistent with the nitrogen abundance that 
corresponds to the galactocentric distance of this \hii\ region
(5.6 kpc), which is 7.0 $\pm$ 0.2, according to Eq. 5 in
\cite{magrini07}.

\begin{figure}
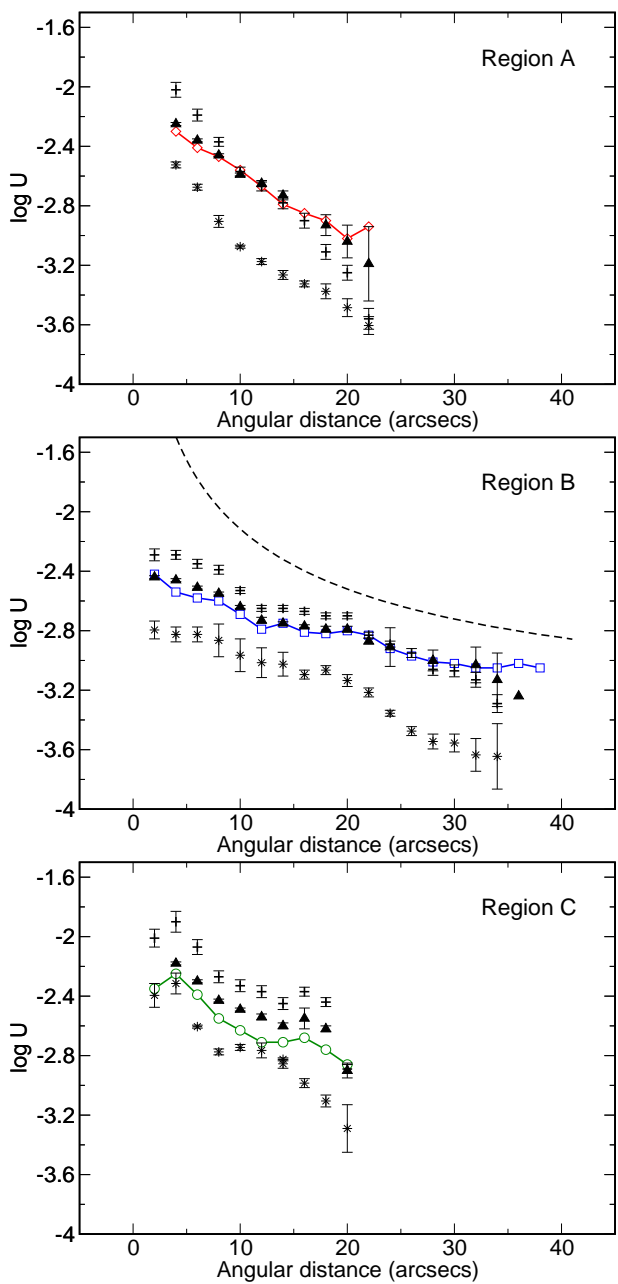

\psfig{figure=logU_A.eps,width=8cm,clip=}
\psfig{figure=logU_B.eps,width=8cm,clip=}
\psfig{figure=logU_C.eps,width=8cm,clip=}
\caption{log $U$ radial profiles for regions A, B, and C, from top to bottom, 
respectively, as derived from the photoionisation models, which are joined by lines 
(solid; thin-shell approach and dashed; middle panel, projected value in the thick-shell approach) and
derived from observations using [\oii]/[\oiii] (up triangles), [\sii]/\hb\ (stars), and
[\oii]/\hb\ (plus symbols) by using the relations with $\log U$ proposed by D\'\i az et al. (2000).
}
\label{logU}
\end{figure}

\subsubsection{Ionisation parameter}

Most of the observed radial profiles in the relative intensities of
the strong emission-lines are associated with an expected variation of
$\log U$ across the nebula that are already pointed out by MI11.  In
Fig. \ref{logU}, we show from top to bottom for the three defined
regions the $\log U$ derived from the models (average value for
thin-shell models and projected U($r$) in the thick-shell).  As can be
seen, the maximum value in all approaches, is reached in the innermost
annuli of the three regions and this value decreases with the distance
to the ionising source by reaching the lowest values in the outermost
annuli of region B.  The cause for this radial decrease is different
depending on the assumed geometry in the models.  In the RB
thick-shell approach, this decrease is mostly owing to the larger
distance from the inner shell of the corresponding annuli to the
ionising source, while this log $U$ decrease is also due to the
variation of the thickness and dust-to-gas ratio across the shell in
the MB thin-shell approach .

The predicted variation of log $U$ in the different model approaches
gives us the possibility to compare it with the values obtained from
different observed strong-line ratios and to test the extent of these
ratios which are defined for RB geometries of integrated observations, in this context.  The resulting $\log U$ obtained from the expressions, as 
proposed by \cite{diaz}, uses three different emission line ratios:
[\oiii]/[\oii], [\sii]/\hb, and [\oii]/\hb\ , which are also shown in
Fig. \ref{logU}.  In the case of the latter two, the used expressions
have a dependence on metallicity, which we fixed at
Z=0.3$\cdot$Z$_{\odot}$. As can be seen, all three indicators for the
three regions lead to a radial decrease of $U$ in the same way of
the models.  The agreement between the $\log U$ derived from the
models and the observational radial profiles derived from the three
strong-line calibrators is good, although it is noticeably
better in the case of [\oiii]/[\oii] and [\oii]/\hb\ , as compared to
[\sii]/\hb, which gives lower values of $\log U$.  This disagreement,
which is less significant in the innermost annuli of region C, cannot
be related in principle to the change in the effective thickness of
the shell, as the offset is approximately the same for all the annuli
of the regions. The disagreement is probably not due to the
dependence of the $\log U$ derived from [\sii]/\hb\ on $Z$, as this
disagreement is not observed for [\oii]/\hb, which also gives an
estimate of $\log U$ with an additional dependence on $Z$. As already pointed out by MI11, this  could be the consequence of a
non-negligible contribution to the [\sii] emission-line flux of the
diffuse gas above the G\hii R (see Fig. 11 in MI11).

In contrast, the projected U($r$) derived from thick-shell models and shown in
the middle panel of Fig. \ref{logU} gives values much higher than those 
obtained from both thin-shell models and emission-line ratios, possibly owing to
the fact that in each point the ionisation structure is not completely covered
in this spherical RB approach.

The $\log U$ variation across the nebula also illustrates the importance of the 
contribution of the regions of low surface brightness to the integrated properties,
as already shown by MI11, for which little variations in
the slit width or position can lead to significant differences in the integrated derived
properties of the nebula.

\begin{figure*}
\begin{minipage}{170mm}
\centerline{
\psfig{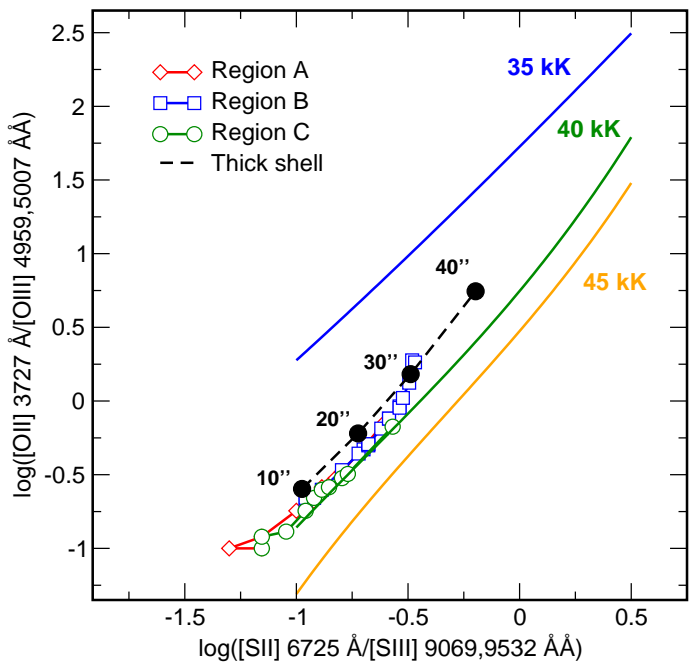}}
\caption{Relation between the ratio [\sii]/[\siii] and [\oii]/[\oiii]. The colour solid lines represent a grid of single star models
for different equivalent effective temperatures for the metallicity of NGC~588. The predictions from thin-shell models (solid line) for different annuli and 
from thick-shell model (dashed line) at different angular distances are also shown.}
\label{Teff}
\end{minipage}
\end{figure*} 

\subsubsection{Equivalent effective temperature}

The equivalent effective temperature depends above all on the shape of
the SED of the ionising source. This can be characterised, for
instance, with the ratio of hydrogen-to-helium ionising photons. In
all the models used to reproduce NGC~588, a single ionising source of
$T_*$ = 41\,000 K was used by adopting the conclusions from the
star-by-star analysis made by \cite{jamet04}, and was kept unchanged
across all the modelled annuli.  Therefore, for this case, it is
important to study to what extent the calibrators based on
strong-emission lines in the optical spectrum can modify the estimation
of $T_*$, as a consequence of the spatial variations across the three
defined regions.

According to \cite{vp88}, the $\eta$ parameter can be used to find the radiation 
hardness. This parameter depends on the relative ionic abundances of
several species whose brightest emission lines can be observed in the optical and near-IR spectrum
(i.e. O$^+$, O$^{2+}$, S$^+$, and S$^{2+}$).
The equivalent to the $\eta$ parameter, which based only on
relative emission-lines if no estimate of the electron temperature
is available, can be expressed as

\begin{equation}
\eta' = \frac{ \textrm{I([O{\sc II}}] \lambda 3727)/\textrm{I([O{\sc III}}] \lambda\lambda 4959,5007)
}{ \textrm{I([S{\sc II}}] \lambda\lambda 6717,6731)/\textrm{I([S{\sc III}}] \lambda\lambda 9069,9532),
}
\end{equation}

\noindent although this emission-line ratio has also an additional dependence
on $Z$ \citep{morisset}. 
A study of the spatial variation of this parameter can
be carried out from the results of the models. Unfortunately, no
spatially resolved measurements of the [\siii] emission line exists.
As in the case of the models made for NGC~595 \citep{mod595}
the $\eta'$ parameter is not uniform across the nebula in NGC~588, according to models.
However, this does not mean that the $T_*$ estimated from this method gives a non-uniform estimate of $T_*$.
As already shown by \cite{sca}, the fits to the grids of photoionisation
models with the same $T_*$ in the [\oii]/[\oiii],[\sii]/[\siii] plane have slopes higher than 1.
Therefore, an estimation of $T_*$ cannot be given only in terms of $\eta'$
but also by means of a direct comparison of the two involved ratios 
and the appropriate sequences of models.
This plot is shown in Fig.\ref{Teff}, where the results of the tailored
models described in this work are compared with the fits to sequences of 
single-star models with different $T_*$ values and the same
metallicity as NGC~588.  
As can be seen, as in the case of NGC~595 \citep{sca}, the models predict that it is 
possible to measure a uniform value for $T_*$ for all the spatial elements, as they
are ionised by the same ionising source, which are independent of the spatial
variations of other physical properties in those regions. This result is found for
the two geometrical assumptions considered in the models.
According to models and lacking of an observational
confirmation in this case, we can thus conclude that despite the log $U$ variations and 
the assumption of different geometries. The $T_*$ estimation from the $\eta'$ method 
is robust even when different spatial elements in an H{\sc ii} regions are used. However, the absolute
scale of this diagnostic must be taken with care, as already shown by \cite{morisset} and
\cite{pmv09}. In this case, the $T_*$ scale is slightly lower than the adopted value for
the SED in these models, which is at 41\,000 K.

\begin{figure}
\psfig{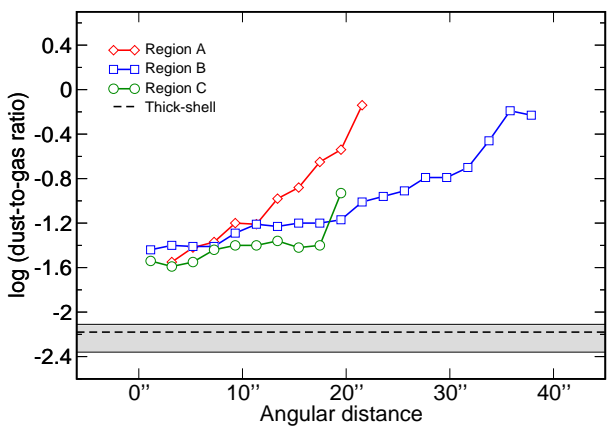}
\caption{Radial variation of the dust-to-gas mass ratio obtained in the MB 
thin-shell models for the three studied
regions (symbols joined by solid lines) and in the RB thick-shell models (dashed line). 
The grey band represents the dust/gas ratio derived
from the integrated IR observations
of the nebula and the mass of H{\sc i}.}
\label{dtg}
\end{figure} 

\begin{figure*}
\begin{minipage}{180mm}
\centerline{
\psfig{figure=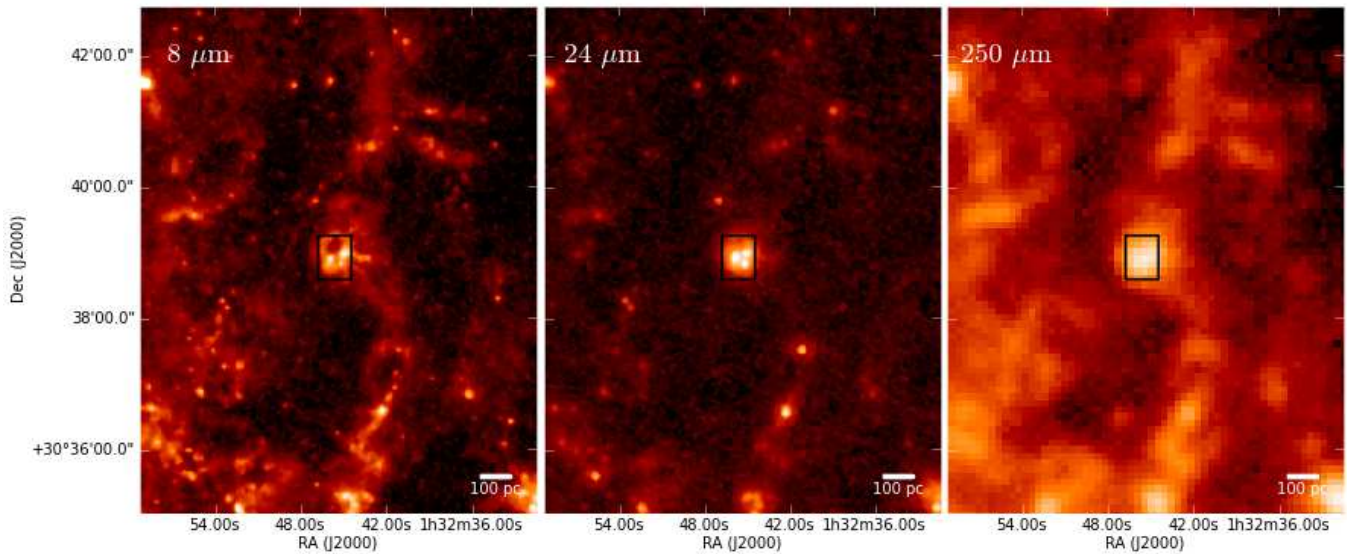,width=18cm,clip=}}
\caption{From left to right, {\em Spitzer} 8~\mi\ and 24~\mi\ and {\em Herschel} 250~\mi\ broadband images of the
area around NGC~588.  The respective beam sizes are 2'', 6'', and 18''.
The black rectangle shows the field of view of the PMAS optical
observations.
A bar showing the physical
scale of 100 pc is plotted at the lower right hand corner of the images. 
}
\label{IRimages}
\end{minipage}
\end{figure*}

\subsection{Dust-to-gas ratio and IR emission}

The dust-to-gas mass ratio is a key parameter in ionised gaseous nebulae
because the relative abundance of dust affects the thermal equilibrium
of the gas.  Unlike the thick-shell approach, where a constant value
of the dust-to-gas mass ratio, agrees with the value derived
from mid-IR observations, is used to match the observations, in the
thin-shell approach the dust-to-gas ratio is introduced as a free
parameter.  As shown in Fig. \ref{dtg}, this ratio has to increase
radially in the models of the three studied regions of the shell to correctly reproduce the observed optical emission-line
ratios, as is the case for high relative intensity of [\oii] and
[\oiii]. Even in the parts of the shell that are closer to the ionising
source, where a lower dust-to-gas ratio is required, the obtained
values are sensibly higher than the dust-to-gas ratio derived from the
integrated observations.  The enhancement of the dust-to-gas ratio
combined with the decrease of the effective thickness for the outer
annuli of the thin-shell have a consequence: the models under
this geometry do not either reproduce the radial profile of the {\em
  Spitzer} 8~\mi-to-24~\mi\ ratio in the outer spatial positions, as
shown in Fig. \ref{IR}.  The radial profiles of the surface brightness
at 8~\mi\ and 24~\mi\ are well traced by these models as fruit of the
geometrical dilution, but the required enhancement of the dust-to-gas
ratio in the outer annuli of the thin-shell models makes the ratio 
decrease in the outer parts. The reason is that the main emitters at
24\mi\ are the smallest and hottest dust grains mixed with the ionised
gas.

In contrast, the increase in the 8~\mi/24~\mi\ ratio in the
thick-shell model is reproduced with a constant dust-to-gas ratio due
to the radial decrease of the dust temperature.  While the dust is hot
and the emission at 24~\mi\ is high in the innermost parts, this dust
is cooler, and the ratio is dominated by the free-free emission at
8~\mi\ in the outer parts according to the models.

Nevertheless, a much larger optical extinction is derived from the Balmer decrement by MI11, 
which would be indicative of different spatial positions for gas and dust.
Other IR sources could be the cause of the high values 
of this ratio in the outer parts of this G\hii R and of the disagreement between
optical and IR extinction.
In Fig. \ref{IRimages}, we show the area around
the region of NGC~588 observed with PMAS, as seen by {\em Spitzer} in 8~\mi\ and 
24~\mi\ and by the {\em Herschel} telescope in 250~\mi\ broadband.
As can be seen, the 8~\mi\ emission of the dust
around the {\hii} region is notably larger than in 24~\mi. This 
emission, which probably comes  from a cloud of diffuse dust above
the plane of the galaxy, is confirmed by inspecting the 250~\mi\ image,
where traces of dust appear at positions that
are not associated with the {\hii} region.  This diffuse emission can thus 
contribute to the 8\mi\ emission at the location of NGC~588 coming mainly from
PAHs.

Regarding visual extinction, no clear spatial trend within the errors 
is found from the Balmer decrement. This is also found in both geometrical
approaches. In the thin-shell models, the decrease in the effective thickness makes the extinction similar in all annuli although dust-to-gas ratio increases with 
distance.

\section{Summary and conclusions}

The two-dimensional optical and mid-IR spatial structure of the 
G\hii R NGC~588 in the disc of M33 was studied by means
of one-dimensional tailor-made photoionisation models. 
This object constitutes an appropriate target to study the spatial interplay
between gas, dust, and stars and, thus, to explore the validity of different methods
used for integrated observations because it is very well characterised 
in observations at several bands.

The observational
source for this study was the work made by MI11,
who describe PMAS - CAHA 3.5 m optical IFU observations and compare them with
{\em Spitzer} 8~\mi\ and 24~\mi\ mid-IR images. For this work,
we also used {\em Spitzer} integrated data at 70~\mi\ and
160~\mi\ to derive the dust-to-gas ratio and {\em Herschel}
250~\mi\ to study the presence of diffuse dust in positions outside
the G\hii R.

To analyse the spatial variation of the observed main emission-line
ratios and the IR emission,
we followed the same procedure as \cite{mod595}
for NGC~595 in which elliptical annuli were defined around the
emission-peak. Although NGC~588 presents a ring-like
morphology as seen in the \ha\ image, it does not have the same
axial symmetry as NGC~595, so different regions were defined to reproduce the different observed patterns. Then, the emission
in the three defined regions was co-added at different angular distances to
the ionising clusters. Finally, the resulting radial profiles for
\hb\ surface brightness and the emission-line ratios of [\oii]/\hb,
[\oiii]/\hb, [\nii]/\ha, and [\sii]/\ha\ were taken to constrain
the models that were made for the different annuli in the three regions.

We adopted two different model strategies to find out the
nature of the observed spatial variations across the field of view in
this GH{\sc ii}R.  On one hand, we considered each spatial element as
having a complete ionisation structure that could be modelled
independently.  This model strategy is an improved version of the
methodology described in \cite{mod595} for NGC~595 and accounts above
all for azimuthal spatial variations of both the observed and the
derived properties in two dimensions.  The resulting geometry in these
models consists of a hollow ellipsoid of revolution, whose MB
thin-shell has most of the emission of the G\hii R.  In this geometry,
the optical emission is fitted by assuming a decreasing thickness and
an increasing dust-to-gas ratio where most of the ionising photons are
absorbed by dust and an additional fraction leaks from the nebula.

On the other hand, we considered that the 2D observed properties of the nebula
are due to the projection of a sphere onto the sky. This was modelled in a single
photoionisation model using the {\sc pyCloudy} code \citep{C3D}.
The resulting geometry in this model consists of a RB 
spherical thick-shell with decreasing particle density.

In both approaches, both the \hb\ surface brightness and the main
optical emission-line ratios were reproduced.  
The models also reproduce the radial variation of the {\em Spitzer}
emission bands at 8~\mi\ and 24~\mi. However, the ratio between
them is underestimated in the outer annuli in the thin-shell geometry
due to the enhancement of the dust-to-gas ratio, which makes the
emission at 24\mi\ to increase there. On the contrary, this ratio is well
reproduced by the thick-shell model. 
We thus conclude that  both projection and azimuthal spatial
variations are possibly present in the properties of the gas and the dust in NGC~588.
However, as the thick-shell approximation achieves a better fit to the mid-IR
observations, the projection effects dominate over the azimuthal ones.
This also shows the importance of fitting both optical and mid-IR spatial
properties to better understand the spatial distribution of the region.

A single-star SED, which is equivalent to the observed stellar clusters as
demonstrated by \cite{jamet04}, was used in the models. This does not
allow a detailed study of the inner ionisation structure in the
G{\hii}R.  However,  the oxygen gas-phase
abundance derived by the models [12+log(O/H) = 8.16, thin-shell, 8.3,
  thick-shell] for both assumptions is uniform across the nebula and agrees with the values
measured by several authors (\citealt{vilchez88}, \citealt{jamet05}), who
use the electron temperature based on optical collisionally excited
emission lines. The conclusions are made by MI11. Under
the assumption of a chemical homogeneity, we reproduced the very high values of \rdostres\ in
the outer annuli of regions A and B only in the thin-shell
models without considering high-velocity
gas-shocks. Therefore, contrary to the rest of the nebula, azimuthal
spatial variations of the geometry and dust-to-gas ratio are possibly
behind the behaviour of these lines at these distances.  According to
both models, the
estimation of $T_*$ from the $\eta'$ parameter is quite uniform across
the nebula despite the high variation of several emission-line
ratios involving both high- and low-excitation emission lines. This indicates that this method is very robust and
independent of the assumed geometry of the models. However, this
result needs to be confirmed with more observations of the spatial
distribution of this parameter.

The MB thin-shell models fit the optical observables by assuming an increasing
dust-to-gas ratio, which go up to much higher values than that derived for the
integrated nebula from IR emission at different bands.  This estimation points 
to a value close to the standard Galactic value  
and was obtained empirically from the total dust mass, 
as found using the dust temperature derived from the ratio 
between 70~\mi\ and 160~\mi, and the total H{\sc i} mass.  
This value is consistent with the dust-to-gas ratio
assumed in the thick-shell approach.
However, the extinction value derived from
this assumption is much lower than the estimates from the optical Balmer
decrement from different authors.
The 8 \mic\ image around the area
of the G\hii R shows structures of the dust, which are not associated
with the {\hii} region and which are confirmed with
the {\em Herschel} image at 250 \mic. The presence of a cloud of
diffuse gas and dust above the plane of the galaxy at the
same position of NGC 588 could also explain the high values of
the 8\mic/24\mic\ ratio.

\section*{Acknowledgements}
This work has been partially supported by projects AYA2007-67965-C03-02, 
AYA2007-67625-C02-02, and AYA2010-21887-C04-01
of the Spanish National Plan for Astronomy and Astrophysics, and by the project
 TIC114  {\em Galaxias y Cosmolog\'\i a} of the
 Junta de Andaluc\'\i a (Spain). This research was also supported 
by a Marie Curie Intra European Fellowship within the 7$^{\rm th}$ 
European Community Framework Programme ERG: PERG08-GA2010-276813.
CM acknowledges support from grant CONACyT CB2010/153985 (M\'exico).

We thank Simon Verley for providing us with the {\em Spitzer} images and for
his help with their reduction and analysis.
We are grateful to the Herschel Open Time Key Project HerM33es for
the 250~\mi\ image used in this paper.
We also thank an anonymous referee whose comments have
largely helped to improve the final manuscript.

This paper uses the plotting package JMAPLOT developed by Jes\'us
Ma\'\i z-Apellaniz (available at
http://jmaiz.iaa.es/iWeb/Software.html).

\bibliographystyle{aa}
\bibliography{aa-n588models}

\begin{thebibliography}{59}
\expandafter\ifx\csname natexlab\endcsname\relax\def\natexlab#1{#1}\fi

\bibitem[{{Alloin} {et~al.}(1979){Alloin}, {Collin-Souffrin}, {Joly}, \&
  {Vigroux}}]{alloin}
{Alloin}, D., {Collin-Souffrin}, S., {Joly}, M., \& {Vigroux}, L. 1979, \aap,
  78, 200

\bibitem[{{Asplund} {et~al.}(2009){Asplund}, {Grevesse}, {Sauval}, \&
  {Scott}}]{asplund}
{Asplund}, M., {Grevesse}, N., {Sauval}, A.~J., \& {Scott}, P. 2009, \araa, 47,
  481

\bibitem[{{Bendo} {et~al.}(2008){Bendo}, {Draine}, {Engelbracht}, {Helou},
  {Thornley}, {Bot}, {Buckalew}, {Calzetti}, {Dale}, {Hollenbach}, {Li}, \&
  {Moustakas}}]{bendo}
{Bendo}, G.~J., {Draine}, B.~T., {Engelbracht}, C.~W., {et~al.} 2008, \mnras,
  389, 629

\bibitem[{{Bluhm} {et~al.}(2003){Bluhm}, {de Boer}, {Marggraf}, {Richter}, \&
  {Wakker}}]{2003A&A...398..983B}
{Bluhm}, H., {de Boer}, K.~S., {Marggraf}, O., {Richter}, P., \& {Wakker},
  B.~P. 2003, \aap, 398, 983

\bibitem[{{Calzetti} {et~al.}(2007){Calzetti}, {Kennicutt}, {Engelbracht},
  {Leitherer}, {Draine}, {Kewley}, {Moustakas}, {Sosey}, {Dale}, {Gordon},
  {Helou}, {Hollenbach}, {Armus}, {Bendo}, {Bot}, {Buckalew}, {Jarrett}, {Li},
  {Meyer}, {Murphy}, {Prescott}, {Regan}, {Rieke}, {Roussel}, {Sheth}, {Smith},
  {Thornley}, \& {Walter}}]{calzetti}
{Calzetti}, D., {Kennicutt}, R.~C., {Engelbracht}, C.~W., {et~al.} 2007, \apj,
  666, 870

\bibitem[{{Castellanos} {et~al.}(2002){Castellanos}, {D{\'{\i}}az}, \&
  {Tenorio-Tagle}}]{marcelo}
{Castellanos}, M., {D{\'{\i}}az}, {\'A}.~I., \& {Tenorio-Tagle}, G. 2002,
  \apjl, 565, L79

\bibitem[{{Cervi{\~n}o} {et~al.}(2003){Cervi{\~n}o}, {Luridiana}, {P{\'e}rez},
  {V{\'{\i}}lchez}, \& {Valls-Gabaud}}]{cervino03}
{Cervi{\~n}o}, M., {Luridiana}, V., {P{\'e}rez}, E., {V{\'{\i}}lchez}, J.~M.,
  \& {Valls-Gabaud}, D. 2003, \aap, 407, 177

\bibitem[{{D{\'{\i}}az} {et~al.}(2000){D{\'{\i}}az}, {Castellanos},
  {Terlevich}, \& {Luisa Garc{\'{\i}}a-Vargas}}]{diaz}
{D{\'{\i}}az}, A.~I., {Castellanos}, M., {Terlevich}, E., \& {Luisa
  Garc{\'{\i}}a-Vargas}, M. 2000, \mnras, 318, 462

\bibitem[{{Dopita} \& {Sutherland}(1996)}]{dopita}
{Dopita}, M.~A. \& {Sutherland}, R.~S. 1996, \apjs, 102, 161

\bibitem[{{Draine} \& {Lee}(1984)}]{1984ApJ...285...89D}
{Draine}, B.~T. \& {Lee}, H.~M. 1984, \apj, 285, 89

\bibitem[{{Ercolano} {et~al.}(2009){Ercolano}, {Bastian}, \&
  {Stasi{\'n}ska}}]{ercolano}
{Ercolano}, B., {Bastian}, N., \& {Stasi{\'n}ska}, G. 2009, \apss, 324, 199

\bibitem[{{Fazio} {et~al.}(2004){Fazio}, {Hora}, {Allen}, {Ashby}, {Barmby},
  {Deutsch}, {Huang}, {Kleiner}, {Marengo}, {Megeath}, {Melnick}, {Pahre},
  {Patten}, {Polizotti}, {Smith}, {Taylor}, {Wang}, {Willner}, {Hoffmann},
  {Pipher}, {Forrest}, {McMurty}, {McCreight}, {McKelvey}, {McMurray}, {Koch},
  {Moseley}, {Arendt}, {Mentzell}, {Marx}, {Losch}, {Mayman}, {Eichhorn},
  {Krebs}, {Jhabvala}, {Gezari}, {Fixsen}, {Flores}, {Shakoorzadeh}, {Jungo},
  {Hakun}, {Workman}, {Karpati}, {Kichak}, {Whitley}, {Mann}, {Tollestrup},
  {Eisenhardt}, {Stern}, {Gorjian}, {Bhattacharya}, {Carey}, {Nelson},
  {Glaccum}, {Lacy}, {Lowrance}, {Laine}, {Reach}, {Stauffer}, {Surace},
  {Wilson}, {Wright}, {Hoffman}, {Domingo}, \& {Cohen}}]{fazio}
{Fazio}, G.~G., {Hora}, J.~L., {Allen}, L.~E., {et~al.} 2004, \apjs, 154, 10

\bibitem[{{Ferland} {et~al.}(1998){Ferland}, {Korista}, {Verner}, {Ferguson},
  {Kingdon}, \& {Verner}}]{ferland}
{Ferland}, G.~J., {Korista}, K.~T., {Verner}, D.~A., {et~al.} 1998, \pasp, 110,
  761

\bibitem[{{Freedman} {et~al.}(1991){Freedman}, {Wilson}, \&
  {Madore}}]{Freedman}
{Freedman}, W.~L., {Wilson}, C.~D., \& {Madore}, B.~F. 1991, \apj, 372, 455

\bibitem[{{Giammanco} {et~al.}(2004){Giammanco}, {Beckman}, {Zurita}, \&
  {Rela{\~n}o}}]{gia04}
{Giammanco}, C., {Beckman}, J.~E., {Zurita}, A., \& {Rela{\~n}o}, M. 2004,
  \aap, 424, 877

\bibitem[{{Gratier} {et~al.}(2010){Gratier}, {Braine}, {Rodriguez-Fernandez},
  {Schuster}, {Kramer}, {Xilouris}, {Tabatabaei}, {Henkel}, {Corbelli},
  {Israel}, {van der Werf}, {Calzetti}, {Garcia-Burillo}, {Sievers}, {Combes},
  {Wiklind}, {Brouillet}, {Herpin}, {Bontemps}, {Aalto}, {Koribalski}, {van der
  Tak}, {Wiedner}, {R{\"o}llig}, \& {Mookerjea}}]{2010A&A...522A...3G}
{Gratier}, P., {Braine}, J., {Rodriguez-Fernandez}, N.~J., {et~al.} 2010, \aap,
  522, A3+

\bibitem[{{Helou} {et~al.}(2004){Helou}, {Roussel}, {Appleton}, {Frayer},
  {Stolovy}, {Storrie-Lombardi}, {Hurt}, {Lowrance}, {Makovoz}, {Masci},
  {Surace}, {Gordon}, {Alonso-Herrero}, {Engelbracht}, {Misselt}, {Rieke},
  {Rieke}, {Willner}, {Pahre}, {Ashby}, {Fazio}, \& {Smith}}]{helou}
{Helou}, G., {Roussel}, H., {Appleton}, P., {et~al.} 2004, \apjs, 154, 253

\bibitem[{{Jamet} \& {Morisset}(2008)}]{jamet08}
{Jamet}, L. \& {Morisset}, C. 2008, \aap, 482, 209

\bibitem[{{Jamet} {et~al.}(2004){Jamet}, {P{\'e}rez}, {Cervi{\~n}o},
  {Stasi{\'n}ska}, {Gonz{\'a}lez Delgado}, \& {V{\'{\i}}lchez}}]{jamet04}
{Jamet}, L., {P{\'e}rez}, E., {Cervi{\~n}o}, M., {et~al.} 2004, \aap, 426, 399

\bibitem[{{Jamet} {et~al.}(2005){Jamet}, {Stasi{\'n}ska}, {P{\'e}rez},
  {Gonz{\'a}lez Delgado}, \& {V{\'{\i}}lchez}}]{jamet05}
{Jamet}, L., {Stasi{\'n}ska}, G., {P{\'e}rez}, E., {Gonz{\'a}lez Delgado},
  R.~M., \& {V{\'{\i}}lchez}, J.~M. 2005, \aap, 444, 723

\bibitem[{{Kramer} {et~al.}(2010){Kramer}, {Buchbender}, {Xilouris}, {Boquien},
  {Braine}, {Calzetti}, {Lord}, {Mookerjea}, {Quintana-Lacaci}, {Rela{\~n}o},
  {Stacey}, {Tabatabaei}, {Verley}, {Aalto}, {Akras}, {Albrecht}, {Anderl},
  {Beck}, {Bertoldi}, {Combes}, {Dumke}, {Garcia-Burillo}, {Gonzalez},
  {Gratier}, {G{\"u}sten}, {Henkel}, {Israel}, {Koribalski}, {Lundgren},
  {Martin-Pintado}, {R{\"o}llig}, {Rosolowsky}, {Schuster}, {Sheth}, {Sievers},
  {Stutzki}, {Tilanus}, {van der Tak}, {van der Werf}, \& {Wiedner}}]{kramer10}
{Kramer}, C., {Buchbender}, C., {Xilouris}, E.~M., {et~al.} 2010, \aap, 518,
  L67

\bibitem[{{Leitherer} {et~al.}(1999){Leitherer}, {Schaerer}, {Goldader},
  {Gonz{\'a}lez Delgado}, {Robert}, {Kune}, {de Mello}, {Devost}, \&
  {Heckman}}]{leitherer}
{Leitherer}, C., {Schaerer}, D., {Goldader}, J.~D., {et~al.} 1999, \apjs, 123,
  3

\bibitem[{{Lisenfeld} {et~al.}(2002){Lisenfeld}, {Israel}, {Stil}, \&
  {Sievers}}]{Lisenfeld}
{Lisenfeld}, U., {Israel}, F.~P., {Stil}, J.~M., \& {Sievers}, A. 2002, \aap,
  382, 860

\bibitem[{{Magrini} {et~al.}(2007){Magrini}, {V{\'{\i}}lchez}, {Mampaso},
  {Corradi}, \& {Leisy}}]{magrini07}
{Magrini}, L., {V{\'{\i}}lchez}, J.~M., {Mampaso}, A., {Corradi}, R.~L.~M., \&
  {Leisy}, P. 2007, \aap, 470, 865

\bibitem[{{Markwardt}(2009)}]{markwardt}
{Markwardt}, C.~B. 2009, in Astronomical Society of the Pacific Conference
  Series, Vol. 411, Astronomical Data Analysis Software and Systems XVIII, ed.
  {D.~A.~Bohlender, D.~Durand, \& P.~Dowler}, 251

\bibitem[{{Mathis} {et~al.}(1977){Mathis}, {Rumpl}, \& {Nordsieck}}]{mrn}
{Mathis}, J.~S., {Rumpl}, W., \& {Nordsieck}, K.~H. 1977, \apj, 217, 425

\bibitem[{{McGaugh}(1991)}]{mcgaugh}
{McGaugh}, S.~S. 1991, \apj, 380, 140

\bibitem[{{Melnick}(1979)}]{melnick79}
{Melnick}, J. 1979, \apj, 228, 112

\bibitem[{{Melnick} {et~al.}(1987){Melnick}, {Moles}, {Terlevich}, \&
  {Garcia-Pelayo}}]{melnick87}
{Melnick}, J., {Moles}, M., {Terlevich}, R., \& {Garcia-Pelayo}, J.-M. 1987,
  \mnras, 226, 849

\bibitem[{{Monreal-Ibero} {et~al.}(2011){Monreal-Ibero}, {Rela{\~n}o},
  {Kehrig}, {P{\'e}rez-Montero}, {V{\'{\i}}lchez}, {Kelz}, {Roth}, \&
  {Streicher}}]{monreal11}
{Monreal-Ibero}, A., {Rela{\~n}o}, M., {Kehrig}, C., {et~al.} 2011, \mnras,
  413, 2242

\bibitem[{{Morisset}(2004)}]{morisset}
{Morisset}, C. 2004, \apj, 601, 858

\bibitem[{{Morisset}(2006)}]{C3D}
{Morisset}, C. 2006, in IAU Symposium, Vol. 234, Planetary Nebulae in our
  Galaxy and Beyond, ed. M.~J. {Barlow} \& R.~H. {M{\'e}ndez}, 467--468

\bibitem[{{Morisset} {et~al.}(2005){Morisset}, {Stasi{\'n}ska}, \&
  {Pe{\~n}a}}]{NEBU3D}
{Morisset}, C., {Stasi{\'n}ska}, G., \& {Pe{\~n}a}, M. 2005, \mnras, 360, 499

\bibitem[{{Mu\~noz-Tu\~n\'on} {et~al.}(1996){Mu\~noz-Tu\~n\'on},
  {Tenorio-Tagle}, {Castaneda}, \& {Terlevich}}]{cmt96}
{Mu\~noz-Tu\~n\'on}, C., {Tenorio-Tagle}, G., {Castaneda}, H.~O., \&
  {Terlevich}, R. 1996, \aj, 112, 1636

\bibitem[{{Pagel} {et~al.}(1979){Pagel}, {Edmunds}, {Blackwell}, {Chun}, \&
  {Smith}}]{pagel79}
{Pagel}, B.~E.~J., {Edmunds}, M.~G., {Blackwell}, D.~E., {Chun}, M.~S., \&
  {Smith}, G. 1979, \mnras, 189, 95

\bibitem[{{Pauldrach} {et~al.}(2001){Pauldrach}, {Hoffmann}, \& {Lennon}}]{wm}
{Pauldrach}, A.~W.~A., {Hoffmann}, T.~L., \& {Lennon}, M. 2001, \aap, 375, 161

\bibitem[{{P{\'e}rez-Montero} \& {Contini}(2009)}]{pmc09}
{P{\'e}rez-Montero}, E. \& {Contini}, T. 2009, \mnras, 398, 949

\bibitem[{{P{\'e}rez-Montero} \& {D{\'{\i}}az}(2003)}]{pmd03}
{P{\'e}rez-Montero}, E. \& {D{\'{\i}}az}, A.~I. 2003, \mnras, 346, 105

\bibitem[{{P{\'e}rez-Montero} \& {D{\'{\i}}az}(2005)}]{pmd05}
{P{\'e}rez-Montero}, E. \& {D{\'{\i}}az}, A.~I. 2005, \mnras, 361, 1063

\bibitem[{{P{\'e}rez-Montero} {et~al.}(2011{\natexlab{a}}){P{\'e}rez-Montero},
  {Rela{\~n}o}, {V{\'{\i}}lchez}, \& {Monreal-Ibero}}]{mod595}
{P{\'e}rez-Montero}, E., {Rela{\~n}o}, M., {V{\'{\i}}lchez}, J.~M., \&
  {Monreal-Ibero}, A. 2011{\natexlab{a}}, \mnras, 412, 675

\bibitem[{{P{\'e}rez-Montero} \& {V{\'{\i}}lchez}(2009)}]{pmv09}
{P{\'e}rez-Montero}, E. \& {V{\'{\i}}lchez}, J.~M. 2009, \mnras, 400, 1721

\bibitem[{{P{\'e}rez-Montero} {et~al.}(2011{\natexlab{b}}){P{\'e}rez-Montero},
  {V{\'{\i}}lchez}, {Rela{\~n}o}, \& {Monreal-Ibero}}]{sca}
{P{\'e}rez-Montero}, E., {V{\'{\i}}lchez}, J.~M., {Rela{\~n}o}, M., \&
  {Monreal-Ibero}, A. 2011{\natexlab{b}}, in Stellar Clusters \& Associations:
  A RIA Workshop on Gaia, 225--228

\bibitem[{{Rela{\~n}o} {et~al.}(2010){Rela{\~n}o}, {Monreal-Ibero},
  {V{\'{\i}}lchez}, \& {Kennicutt}}]{relano10}
{Rela{\~n}o}, M., {Monreal-Ibero}, A., {V{\'{\i}}lchez}, J.~M., \& {Kennicutt},
  R.~C. 2010, \mnras, 402, 1635

\bibitem[{Rela{\~n}o \& Kennicutt(2009)}]{relano09}
Rela{\~n}o, M. \& Kennicutt, R.~C. 2009, ApJ, 699, 1125

\bibitem[{{Rieke} {et~al.}(2004){Rieke}, {Young}, {Engelbracht}, {Kelly},
  {Low}, {Haller}, {Beeman}, {Gordon}, {Stansberry}, {Misselt}, {Cadien},
  {Morrison}, {Rivlis}, {Latter}, {Noriega-Crespo}, {Padgett}, {Stapelfeldt},
  {Hines}, {Egami}, {Muzerolle}, {Alonso-Herrero}, {Blaylock}, {Dole}, {Hinz},
  {Le Floc'h}, {Papovich}, {P{\'e}rez-Gonz{\'a}lez}, {Smith}, {Su}, {Bennett},
  {Frayer}, {Henderson}, {Lu}, {Masci}, {Pesenson}, {Rebull}, {Rho}, {Keene},
  {Stolovy}, {Wachter}, {Wheaton}, {Werner}, \& {Richards}}]{rieke}
{Rieke}, G.~H., {Young}, E.~T., {Engelbracht}, C.~W., {et~al.} 2004, \apjs,
  154, 25

\bibitem[{{Rosolowsky} \& {Simon}(2008)}]{RS2008}
{Rosolowsky}, E. \& {Simon}, J.~D. 2008, \apj, 675, 1213

\bibitem[{{Roth} {et~al.}(2010){Roth}, {Fechner}, {Wolter}, {Sandin}, {Kelz},
  {Bauer}, {Popow}, {Monreal-Ibero}, {Kehrig}, \& {Streicher}}]{Roth10}
{Roth}, M.~M., {Fechner}, T., {Wolter}, D., {et~al.} 2010, in Society of
  Photo-Optical Instrumentation Engineers (SPIE) Conference Series, Vol. 7742,
  Society of Photo-Optical Instrumentation Engineers (SPIE) Conference Series

\bibitem[{{Roth} {et~al.}(2005){Roth}, {Kelz}, {Fechner}, {Hahn}, {Bauer},
  {Becker}, {B{\"o}hm}, {Christensen}, {Dionies}, {Paschke}, {Popow}, {Wolter},
  {Schmoll}, {Laux}, \& {Altmann}}]{roth}
{Roth}, M.~M., {Kelz}, A., {Fechner}, T., {et~al.} 2005, \pasp, 117, 620

\bibitem[{{Rubin} {et~al.}(2008){Rubin}, {Simpson}, {Colgan}, {Dufour},
  {Brunner}, {McNabb}, {Pauldrach}, {Erickson}, {Haas}, \& {Citron}}]{rubin08}
{Rubin}, R.~H., {Simpson}, J.~P., {Colgan}, S.~W.~J., {et~al.} 2008, \mnras,
  387, 45

\bibitem[{{Storchi-Bergmann} {et~al.}(1998){Storchi-Bergmann}, {Schmitt},
  {Calzetti}, \& {Kinney}}]{storchi}
{Storchi-Bergmann}, T., {Schmitt}, H.~R., {Calzetti}, D., \& {Kinney}, A.~L.
  1998, \aj, 115, 909

\bibitem[{Tabatabaei {et~al.}(2007)Tabatabaei, Beck, Kr{\"u}gel,
  {et~al.}}]{Tabatabaei:2007p664}
Tabatabaei, F.~S., Beck, R., Kr{\"u}gel, E., {et~al.} 2007, A{\&}A, 475, 133

\bibitem[{{van den Bergh}(2000)}]{vandenBergh}
{van den Bergh}, S. 2000, {The Galaxies of the Local Group} (Cambridge)

\bibitem[{{Verley} {et~al.}(2007){Verley}, {Hunt}, {Corbelli}, \&
  {Giovanardi}}]{verley}
{Verley}, S., {Hunt}, L.~K., {Corbelli}, E., \& {Giovanardi}, C. 2007, \aap,
  476, 1161

\bibitem[{Verley {et~al.}(2007)Verley, Hunt, Corbelli,
  {et~al.}}]{Verley:2007p574}
Verley, S., Hunt, L.~K., Corbelli, E., {et~al.} 2007, A{\&}A, 476, 1161

\bibitem[{{Verley} {et~al.}(2010){Verley}, {Rela{\~n}o}, {Kramer}, {Xilouris},
  {Boquien}, {Calzetti}, {Combes}, {Buchbender}, {Braine}, {Quintana-Lacaci},
  {Tabatabaei}, {Lord}, {Israel}, {Stacey}, \& {van der Werf}}]{verley10}
{Verley}, S., {Rela{\~n}o}, M., {Kramer}, C., {et~al.} 2010, \aap, 518, L68

\bibitem[{{Viallefond} \& {Goss}(1986)}]{vg86}
{Viallefond}, F. \& {Goss}, W.~M. 1986, \aap, 154, 357

\bibitem[{{V{\'{\i}}lchez} \& {Pagel}(1988)}]{vp88}
{V{\'{\i}}lchez}, J.~M. \& {Pagel}, B.~E.~J. 1988, \mnras, 231, 257

\bibitem[{{V\'{\i}lchez} {et~al.}(1988){V\'{\i}lchez}, {Pagel}, {Diaz},
  {Terlevich}, \& {Edmunds}}]{vilchez88}
{V\'{\i}lchez}, J.~M., {Pagel}, B.~E.~J., {Diaz}, A.~I., {Terlevich}, E., \&
  {Edmunds}, M.~G. 1988, \mnras, 235, 633

\bibitem[{{Xilouris} {et~al.}(2012){Xilouris}, {Tabatabaei}, {Boquien},
  {Kramer}, {Buchbender}, {Bertoldi}, {Anderl}, {Braine}, {Verley},
  {Rela{\~n}o}, {Quintana-Lacaci}, {Akras}, {Beck}, {Calzetti}, {Combes},
  {Gonzalez}, {Gratier}, {Henkel}, {Israel}, {Koribalski}, {Lord}, {Mookerjea},
  {Rosolowsky}, {Stacey}, {Tilanus}, {van der Tak}, \& {van der
  Werf}}]{xilouris}
{Xilouris}, E.~M., {Tabatabaei}, F.~S., {Boquien}, M., {et~al.} 2012, \aap,
  543, A74

\end{thebibliography}

\end{document}